\documentclass[review,3p]{elsarticle}
\usepackage{amsmath,amssymb,bm}
\usepackage{graphicx,booktabs,tabularx,multirow}
\usepackage{algorithm,algorithmic}
\usepackage[hidelinks]{hyperref}
\usepackage[section]{placeins}
\usepackage{microtype}
\usepackage{enumitem}

\newcommand{\Kn}{\mathrm{Kn}}
\newcommand{\dd}{\,\mathrm{d}}
\newcommand{\syn}{\mathrm{syn}}
\newcommand{\kin}{\mathrm{kin}}
\newcommand{\FR}{\mathrm{FR}}
\newcommand{\GK}{\mathrm{GK}}

\newtheorem{proposition}{Proposition}
\newtheorem{corollary}{Corollary}
\newtheorem{remark}{Remark}
\newenvironment{proofof}{\par\noindent\textit{Proof.}\ }{\hfill$\square$\par\medskip}

\journal{Journal of Computational Physics}

\begin{document}
\begin{frontmatter}
\title{Proper generalized decomposition with synthetic preconditioning for multiscale parametric phonon transport}

\author[label1]{Jia Liu}
\author[label1,label3]{Wei Su\corref{cor1}}
\ead{weisu@ust.hk}
\cortext[cor1]{Corresponding author}

\affiliation[label1]{
            organization={Division of Emerging Interdisciplinary Areas},
            addressline={The Hong Kong University of Science and Technology},
            city={Hong Kong S.A.R}}
\affiliation[label3]{
            organization={Department of Mathematics},
            addressline={The Hong Kong University of Science and Technology},
            city={Hong Kong S.A.R}}

\begin{abstract}
Deterministic solution of the phonon Boltzmann transport equation is costly for two reasons: the distribution depends on high-dimensional phase space, and source iteration converges slowly when scattering is frequent. We address both for the steady linearized gray Callaway model in two dimensions by combining proper generalized decomposition (PGD) with the general synthetic iterative scheme (GSIS). The distribution is a sum of products of a space--azimuth factor and a polar factor. This splitting is intrinsically low-rank: along each characteristic the polar angle only stretches the in-plane optical depth, so the rank is bounded, uniformly in the Knudsen numbers, by the numerical rank of the kernel $\exp(-\tau/\sin\phi)$. PGD--GSIS updates the factors against a macroscopic source frozen within each alternating step and refreshed by the GSIS moment equations. A Fourier analysis shows that pure PGD, although implicit in the polar factor, moves macroscopic information only through lagged spatial sweeps and stalls like source iteration, whereas the synthetic refresh sees only third-order moments of the representation error. On an isothermal cavity, cross-plane conduction and phonon Poiseuille flow, PGD--GSIS reproduces full-order solutions at ranks 3--14, with heat-flux errors below $6\times10^{-3}$ from the diffusive to the ballistic regime and memory independent of the polar resolution. Treating the inverse Knudsen numbers as coordinates of the polar factor yields one spatial basis over four decades of both; a joint update of all polar factors, which costs only small nodal linear systems, is required for accuracy at large Knudsen numbers.
\end{abstract}

\end{frontmatter}

\section{Introduction}\label{sec:introduction}
Heat conduction departs from Fourier's law when device dimensions become comparable to phonon mean free paths~\cite{majumdar1993microscale,joshi1993transient,chen2021non}. The phonon Boltzmann transport equation (BTE) describes this regime by following the phonon distribution in physical and velocity space. Combined with the Callaway relaxation model~\cite{callaway1959model,holland1963analysis}, which distinguishes momentum-destroying resistive (R) scattering from momentum-conserving normal (N) scattering, a single equation covers ballistic, diffusive and hydrodynamic transport, including phonon Poiseuille flow~\cite{guyer1966solution,guyer1966thermal,guo2015phonon,cepellotti2015phonon,lee2015hydrodynamic}. This generality has a price: deterministic solutions of the BTE are expensive, and design studies or uncertainty quantification, which repeat them over ranges of scattering rates, multiply that expense.

The cost has two distinct sources. The first is the size of the state. Even for a gray model in two spatial dimensions, a discrete-ordinate method stores the distribution at every spatial degree of freedom and every ordinate, so storage grows with the product of the spatial and angular resolutions. Monte Carlo methods avoid this storage at the price of statistical noise, which variance-reduced formulations reduce substantially~\cite{peraud2011efficient}. The second source is slow convergence. Source iteration alternates a transport sweep with an update of the lagged scattering source and propagates information by about one mean free path per sweep, so its error-reduction factor approaches one as collisions become frequent~\cite{adams2002fast}. Diffusion synthetic acceleration~\cite{alcouffe1977diffusion}, multiscale kinetic schemes~\cite{xu2010unified,zhu2017implicit,zhang2019discrete,luo2017discrete,liu2025unified} and the general synthetic iterative scheme (GSIS)~\cite{Su2020JCP,Su2020SIAM,liu2022fast} remove this stiffness. Between kinetic sweeps, GSIS solves macroscopic moment equations whose higher-order terms are evaluated from the current kinetic state. For the phonon Callaway model it reaches the steady state in tens to a few hundred iterations over the whole range of Knudsen numbers, where source iteration may need thousands~\cite{liu2022fast}. It does not, however, reduce the size of the state.

Low-rank methods target this first source of cost. Proper generalized decomposition (PGD)~\cite{ammar2006new,chinesta2011overview,nouy2010priori,chinesta2013proper} builds a sum of separated products directly from the governing equation, without precomputed snapshots: a greedy procedure adds one product at a time, and alternating Galerkin updates determine its factors. Model parameters can be treated as additional coordinates, so that one representation covers a whole parameter domain~\cite{sevilla2020solution,tsiolakis2022parametric}. PGD has been applied to neutron transport~\cite{dominesey2023reduced} and to rarefied gas flow~\cite{su2025reduced}; for the parametrized Shakhov model, adding synthetic equations to the PGD iteration makes the reduced solver recover the Navier--Stokes limit~\cite{yin2026asymptotic}. Reduced-basis methods~\cite{peng2022reduced} instead build low-dimensional spaces from full-order solutions, and dynamical low-rank approximation evolves factorized solutions in time~\cite{einkemmer2018low,peng2020low,hu2022adaptive,kusch2021dynamical}; its asymptotic-preserving variant based on a micro--macro decomposition~\cite{einkemmer2021collision} shares our aim of obtaining the macroscopic limit independently of the rank.

Combining PGD with synthetic acceleration raises two questions. The first is which splitting to use. In transport problems the upwind direction ties space to angle, and a splitting that separates space from direction altogether loses this structure, because the spatial factor then sees only direction-averaged streaming coefficients. The second is how the separated iteration converges. A separated representation does not change how macroscopic information propagates. If the scattering source is taken from the moments of the current representation, the iteration remains a source iteration in disguise: even when this coupling is treated implicitly in the polar factor, macroscopic information crosses the domain only through transport sweeps of the spatial factor, and enrichment adds modes to an unconverged state. This stall occurs precisely where the solution is most compressible: the leading Fourier and Guyer--Krumhansl equilibria have rank at most two in the splitting used below. The synthetic step of GSIS, in contrast, needs only a few angular moments of the kinetic state, which a separated representation supplies cheaply. Whether the truncation and alternation errors of PGD interfere with that step requires analysis of the PGD iteration itself and of the matching between reconstructed and synthetic fields.

In this paper we combine PGD with GSIS for the steady linearized gray Callaway model on two-dimensional domains. Our contributions are the following.
\begin{enumerate}[leftmargin=1.6em,itemsep=2pt]
\item \emph{A splitting with a Knudsen-uniform rank bound.} The distribution is separated into a space--azimuth factor, which carries the upwind structure and is computed by standard discontinuous Galerkin (DG) sweeps, and a polar factor. Because the polar angle only stretches the in-plane optical depth, the separation rank is bounded, uniformly in the Knudsen numbers, by the numerical rank of the kernel $\mathrm{e}^{-\tau/\sin\phi}$; any $P_N$ angular state has rank at most $N+1$ (Section~\ref{sec:pgd}).
\item \emph{Synthetic coupling to the separated state.} PGD--GSIS freezes the macroscopic source within each alternating pair and refreshes it by the synthetic equations, discretized by a hybridizable DG (HDG) method. Their high-order loads, including DG face terms, and the half-space wall closure are assembled from only four polar moments of each mode (Section~\ref{sec:maps} and \ref{app:hdg}).
\item \emph{A convergence analysis of pure PGD and PGD--GSIS.} Pure PGD, although implicit in the polar factor, moves macroscopic information only through lagged spatial sweeps and inherits the stall of source iteration. Each PGD--GSIS step splits into the full-order GSIS error propagation~\cite{liu2022fast} and the image of the representation error, which the synthetic map reads only through third-order angular moments. In the ballistic limit the roles reverse (Section~\ref{sec:cis_fourier}).
\item \emph{An analysis of the asymptotic limits.} A formal expansion of the iteration identifies the required polar space; explicit conditions on the polar space, the projected residual and the source increment then preserve the leading Fourier flux and the viscous GK balance, and in the ballistic limit the vanishing collision rate cancels the $\mathcal O(\Kn_C)$ synthetic gain (Section~\ref{sec:ap} and \ref{app:limits}).
\item \emph{A parametric extension}, in which the polar factor also depends on both inverse Knudsen numbers, so that one spatial basis covers four decades of each. A joint update of all polar factors, one small linear system per polar and parameter node, is required for accuracy at large Knudsen numbers. We identify what sets the rank and where the error concentrates (Section~\ref{sec:parametric}).
\end{enumerate}
Section~\ref{sec:model} introduces the model, its macroscopic structure and the test problems. Sections~\ref{sec:pgd}--\ref{sec:parametric} develop and analyze the separated representation, the alternating iteration and the parametric extension. Section~\ref{sec:results} reports results for an isothermal cavity, cross-plane conduction and phonon Poiseuille flow, and Section~\ref{sec:conclusions} concludes. The appendices contain the asymptotic analysis, the Fourier form of the iteration, and the DG and HDG discretizations.

\section{Kinetic model and its macroscopic structure}\label{sec:model}
\subsection{Linearized gray Callaway equation}
Positions $\bm x=(x,y)$ lie in a two-dimensional domain $\mathcal D$ with no variation in the third direction, while phonon directions are three-dimensional,
\begin{equation}
 \bm\Omega=(\sin\phi\cos\theta,\ \sin\phi\sin\theta,\ \cos\phi)=(\sin\phi\,\bm s(\theta),\cos\phi),
 \qquad \dd\Omega=\sin\phi\dd\phi\dd\theta,
 \label{eq:direction}
\end{equation}
with azimuth $\theta\in[0,2\pi)$ and polar angle $\phi\in[0,\pi]$. Material properties are gray and constant, temperature deviations from a reference $T_0$ are small, and there is no volumetric heat source. Let $\tilde I(\tilde{\bm x},\bm\Omega)$ be the deviational energy density per unit solid angle, integrated over frequency and polarization, so that $\int\tilde I\dd\Omega=C_v(\tilde T-T_0)$ and $\tilde{\bm q}=v_g\int\bm\Omega\tilde I\dd\Omega$. The steady linearized Callaway model~\cite{callaway1959model,holland1963analysis,liu2022fast} reads
\begin{equation}
 v_g\bm\Omega\cdot\widetilde\nabla\tilde I
 =\frac{\tilde I_R^{\rm eq}-\tilde I}{\tau_R}
  +\frac{\tilde I_N^{\rm eq}-\tilde I}{\tau_N},\qquad
 \tilde I_R^{\rm eq}=\frac{C_v(\tilde T-T_0)}{4\pi},\quad
 \tilde I_N^{\rm eq}=\tilde I_R^{\rm eq}
       +\frac{3\bm\Omega\cdot\tilde{\bm q}}{4\pi v_g}.
 \label{eq:bte_dim}
\end{equation}
With a length $L$, a temperature scale $\Delta T$, $\bm x=\tilde{\bm x}/L$, $I=\tilde I/(C_v\Delta T)$, $T=(\tilde T-T_0)/\Delta T$, $\bm q=\tilde{\bm q}/(C_vv_g\Delta T)$, and the Knudsen numbers $\Kn_{R,N}=v_g\tau_{R,N}/L$, define the inverse Knudsen numbers $\delta_R=\Kn_R^{-1}$, $\delta_N=\Kn_N^{-1}$ and $\delta_C=\delta_R+\delta_N=\Kn_C^{-1}$. Then
\begin{equation}
 \bm\Omega\cdot\nabla I+\delta_C I
 =\frac{\delta_C}{4\pi}T+\frac{3\delta_N}{4\pi}\bm\Omega\cdot\bm q,
 \qquad T=\int I\dd\Omega,\quad \bm q=\int\bm\Omega I\dd\Omega.
 \label{eq:bte}
\end{equation}
The right-hand side is $\delta_C$ times the displaced equilibrium $(T+3\chi\,\bm\Omega\cdot\bm q)/(4\pi)$, where $\chi=\delta_N/\delta_C\in[0,1)$ is the fraction of collisions that conserve momentum. The out-of-plane flux vanishes by the symmetry $\phi\mapsto\pi-\phi$ of the data, so $\bm q=(q_x,q_y)$.

\subsection{Moment equations and the synthetic system}
With the trace-free moments
\begin{equation}
 N_{ij}=\int\Bigl(\Omega_i\Omega_j-\frac{\delta_{ij}}{3}\Bigr)I\dd\Omega,\qquad
 M_{ijk}=\int K_{ijk}(\bm\Omega) I\dd\Omega,
 \label{eq:moments}
\end{equation}
where $K_{ijk}=\Omega_i\Omega_j\Omega_k-(\Omega_i\delta_{jk}+\Omega_j\delta_{ik}+\Omega_k\delta_{ij})/5$ and $\delta_{ij}$ is the Kronecker symbol, the $1$, $\Omega_i$ and $\Omega_i\Omega_j-\delta_{ij}/3$ moments of Eq.~\eqref{eq:bte} are, exactly (repeated indices summed, $\partial_z=0$),
\begin{equation}
 \partial_k q_k=0,\qquad
 \frac13\partial_iT+\partial_kN_{ik}=-\delta_Rq_i,\qquad
 \frac25\partial_{\langle j}q_{i\rangle}+\partial_kM_{ijk}=-\delta_CN_{ij},
 \label{eq:moment_system}
\end{equation}
with $\partial_{\langle j}q_{i\rangle}=(\partial_jq_i+\partial_iq_j)/2-\delta_{ij}\partial_kq_k/3$. Solving the last relation for $N_{ij}$ and substituting it into the second, with $H_{ij}=\partial_kM_{ijk}$, gives
\begin{equation}
 \frac13\nabla T+\delta_R\bm q
 -\frac{\Kn_C}{5}\Bigl[\nabla^2\bm q+\frac13\nabla(\nabla\cdot\bm q)\Bigr]
 =\Kn_C\,\nabla\cdot\bm H,\qquad
 \nabla\cdot\bm q=0.
 \label{eq:synthetic}
\end{equation}
No closure has been made: if $\bm H$ is evaluated from a kinetic state, Eq.~\eqref{eq:synthetic} is an exact rearrangement of Eq.~\eqref{eq:moment_system}. This is the macroscopic (synthetic) system of GSIS~\cite{Su2020SIAM,liu2022fast}. It has the form of the Brinkman equations, with $\bm q$ in the role of velocity, $T/3$ of pressure, $\delta_R$ of Darcy friction and $\Kn_C/5$ of viscosity. It couples the whole domain in one solve, and the kinetic state enters only through $\bm H$, with the explicit factor $\Kn_C$. Setting $\bm H=0$ and multiplying Eq.~\eqref{eq:synthetic} by $\Kn_R$ gives the Guyer--Krumhansl balance~\cite{guyer1966solution,guyer1966thermal}
\begin{equation}
 \bm q-\ell_\GK^2\Bigl[\nabla^2\bm q+\frac13\nabla(\nabla\cdot\bm q)\Bigr]=-\frac{\Kn_R}{3}\nabla T,
 \qquad \ell_\GK^2=\frac{\Kn_R\Kn_C}{5},
 \label{eq:gk}
\end{equation}
for smooth bulk fields with $\Kn_C\ll L_g$, where $L_g$ is the characteristic spatial variation length (\ref{app:limits}); at $\Kn_C\gtrsim L_g$, kinetic effects must be retained. The ratio $\ell_\GK/L_g$ measures viscous effects relative to resistive relaxation: small ratios give Fourier-like transport, while ratios of order one or larger allow hydrodynamic behavior where the flux varies spatially. Normal-scattering dominance alone does not fix this ratio: at fixed $\Kn_R$, $\Kn_N\to0$ gives $\ell_\GK\to0$ and the strict bulk limit is Fourier conduction.

\subsection{Boundary conditions}\label{sec:bc}
Every boundary prescribes the incoming half-range $\bm\Omega\cdot\bm n<0$, with $\bm n$ the outward normal, and leaves the outgoing half-range to the interior solution. A \emph{thermalizing wall} at temperature $T_b$ emits $I=T_b/(4\pi)$. A \emph{diffuse adiabatic wall} emits an isotropic intensity whose level cancels the outgoing normal energy flux,
\begin{equation}
 I_b=\frac{\int_{\bm\Omega\cdot\bm n>0}(\bm\Omega\cdot\bm n)I\dd\Omega}{-\int_{\bm\Omega\cdot\bm n<0}(\bm\Omega\cdot\bm n)\dd\Omega};
 \label{eq:adiabatic}
\end{equation}
numerator and denominator are evaluated with the same discrete ordinates, so the discrete wall flux vanishes exactly. On \emph{periodic} faces paired by a translation $\bm\Delta$, the incoming intensity is the partner's intensity at the same ordinate plus $\Delta T_b/(4\pi)$; $\Delta T_b\neq0$ imposes a mean temperature drop along $\bm\Delta$. All three are isotropic in the incoming half-range or copy the partner's state; none introduces a polar dependence, a fact used in Section~\ref{sec:characteristics}. Without a thermalizing wall the solution is defined up to $I\mapsto I+c/(4\pi)$; the mean temperature is then fixed by Eq.~\eqref{eq:gauge}, and the constant equilibrium is stored separately from the separated representation.

\subsection{Test problems}\label{sec:tests}
\begin{figure}[tb]
\centering
\includegraphics[width=0.76\textwidth]{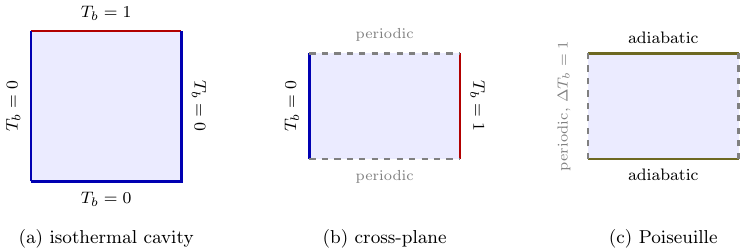}
\caption{Test problems (not to scale). Solid edges are thermalizing (red hot, blue cold) or diffusely adiabatic (olive); dashed edges are periodic. The channels have height $H=0.2$.}
\label{fig:cases}
\end{figure}

The three problems of Fig.~\ref{fig:cases} are the gray-Callaway benchmarks used to validate full-order GSIS~\cite{liu2022fast}; each isolates a different macroscopic mechanism.

\emph{Isothermal cavity.} The unit square has its top wall at $T_b=1$ and the other three walls at $T_b=0$. In the Fourier regime the bulk temperature is harmonic; in the hydrodynamic regime the bulk becomes nearly isothermal and the temperature variation concentrates in wall layers.

\emph{Cross-plane conduction.} The channel $[0,1]\times[0,H]$, $H=0.2$, has thermalizing walls at $x=0$ ($T_b=0$) and $x=1$ ($T_b=1$); the sides $y=0,H$ are periodic. The solution depends on $x$ only, so $\nabla\cdot\bm q=0$ forces a uniform $q_x$ and the viscous term in Eq.~\eqref{eq:gk} vanishes identically. The bulk gradient is then $\partial_xT=-3\delta_Rq_x$: linear profiles in the Fourier regime, and a bulk plateau with large wall temperature jumps whenever $\delta_R$ is small, whether scattering is hydrodynamic or absent.

\emph{Phonon Poiseuille flow.} The same channel has diffusely adiabatic walls at $y=0,H$, and its ends $x=0,1$ are periodic with $\Delta T_b=1$, which imposes a unit mean temperature gradient along $x$. Here the viscous term is active: Eq.~\eqref{eq:gk} gives $q_x(y)\propto1-\cosh[(y-H/2)/\ell_\GK]/\cosh[H/(2\ell_\GK)]$ plus wall slip. The profile is flat with thin wall layers when $\ell_\GK\ll H$ and parabolic when $\ell_\GK\gg H$, with center value $\approx5H^2/(24\Kn_C)$, independent of $\Kn_R$: momentum is then limited by wall friction rather than by resistive scattering.

\section{Separated representation and its rank}\label{sec:pgd}
\subsection{Space--azimuth by polar splitting}
For fixed $(\Kn_R,\Kn_N)$ the distribution is approximated by
\begin{equation}
 I_m(\bm x,\theta,\phi)=\sum_{r=1}^mX_r(\bm x,\theta)\,\Phi_r(\phi).
 \label{eq:separated}
\end{equation}
The in-plane streaming operator is $\sin\phi\,\bm s(\theta)\cdot\nabla$ with $\sin\phi\geq0$, so on every face the inflow/outflow decision depends on $\theta$ alone. The space--azimuth factor therefore carries the whole upwind structure of the two-dimensional geometry and is computed by ordinary DG transport sweeps, one per azimuth; the polar factor never needs upwinding. In three spatial dimensions the term $\cos\phi\,\partial_z$ would make the upwind direction depend on $\phi$, and this property would be lost.

The macroscopic fields follow from azimuthal contractions of $X_r$ and polar moments of $\Phi_r$:
\begin{equation}
 Y_r=\int_0^{2\pi}X_r\dd\theta,\quad
 \bm Z_r=\int_0^{2\pi}\bm sX_r\dd\theta,\qquad
 \mathsf M_k(\Phi)=\int_0^\pi\Phi\sin^k\phi\dd\phi,
 \label{eq:macro_coeff}
\end{equation}
with $\mathsf T_r=\mathsf M_1(\Phi_r)$ and $\mathsf Q_r=\mathsf M_2(\Phi_r)$, so that
\begin{equation}
 T_m^{\kin}=\sum_{r=1}^mY_r\mathsf T_r,\qquad
 \bm q_m^{\kin}=\sum_{r=1}^m\bm Z_r\mathsf Q_r.
 \label{eq:macro_reconstruction}
\end{equation}
The factor equations below use the product measure $\dd\bm x\dd\theta\dd\phi$; the powers of $\sin\phi$ in Eq.~\eqref{eq:macro_coeff} come from $\dd\Omega$ and from the in-plane velocity. Higher moments need only higher polar moments: the in-plane components of $N_{ij}$ involve $\mathsf M_1$ and $\mathsf M_3$, those of $M_{ijk}$ involve $\mathsf M_2$ and $\mathsf M_4$. The entire macroscopic coupling of Section~\ref{sec:gsis} therefore sees each polar factor through four numbers.

\subsection{Low-rank structure and a Knudsen-uniform bound}\label{sec:characteristics}
The rank of the splitting is set by how $\phi$ enters the transport operator. For a given macroscopic source, integrate Eq.~\eqref{eq:bte} along the characteristic through $\bm x$ in direction $\bm\Omega$, parameterized by the \emph{in-plane} distance $t$, so that the path length is $t/\sin\phi$:
\begin{equation}
 I(\bm x,\theta,\phi)=I_b\,\mathrm{e}^{-\delta_Cd/\sin\phi}
 +\int_0^{d}\Bigl[A(\bm x-t\bm s)+\sin\phi\,B(\bm x-t\bm s,\theta)\Bigr]\mathrm{e}^{-\delta_Ct/\sin\phi}\,\frac{\dd t}{\sin\phi},
 \label{eq:characteristic}
\end{equation}
with $A=\delta_CT/(4\pi)$, $B=3\delta_N\bm s\cdot\bm q/(4\pi)$, and $d(\bm x,\theta)$ the in-plane distance to the upstream boundary point $\bm x_b$. The geometry depends on $(\bm x,\theta)$ only, and the boundary data of Section~\ref{sec:bc} do not depend on $\phi$; periodic faces continue the ray through the periodic image. Integrating the $A$-term by parts in $t$ removes the factor $1/\sin\phi$:
\begin{equation}
 I(\bm x,\theta,\phi)=\int_{[0,\infty)}\kappa_\tau(\phi)\dd m_{\bm x,\theta}(\tau),
 \qquad \kappa_\tau(\phi)=\mathrm{e}^{-\tau/\sin\phi},
 \label{eq:kernel_rep}
\end{equation}
where $\tau=\delta_Ct$ is the in-plane optical depth. The measure $m_{\bm x,\theta}$ has point masses $T(\bm x)/(4\pi)$ at $\tau=0$ and $I_b-T(\bm x_b)/(4\pi)$ at $\tau=\delta_Cd$, and density $(\partial_\tau T+3\chi\,\bm s\cdot\bm q)/(4\pi)$ per unit optical depth. All polar dependence is thus carried by the Laplace-type kernel family $\{\kappa_\tau\}_{\tau\geq0}$. Let $p_\eta$ be the smallest $p$ such that $\|\kappa_\tau-\sum_{k=1}^pu_k(\tau)v_k\|\leq\eta\|\kappa_0\|$ for all $\tau\geq0$, in the $L^2(0,\pi)$ norm.

\begin{proposition}[Knudsen-uniform separation rank]\label{prop:rank}
Let $T$ and $\bm q$ be bounded, with $T$ of bounded variation along rays, and let $I$ solve Eq.~\eqref{eq:bte} with this gain term and the boundary conditions of Section~\ref{sec:bc}. For every $\eta\in(0,1)$ there are functions $X_1,\dots,X_{p_\eta}$ of $(\bm x,\theta)$ such that, at every $(\bm x,\theta)$,
\begin{equation}
 \Bigl\|I(\bm x,\theta,\cdot)-\sum_{k=1}^{p_\eta}X_k(\bm x,\theta)\,v_k\Bigr\|
 \leq2\eta\,(1+\ln\eta^{-1})\,\|\kappa_0\|\Bigl[|I_b|+\frac{\|T\|_\infty+V(T)+3\chi\|\bm q\|_\infty}{4\pi}\Bigr],
 \label{eq:rank_error}
\end{equation}
where $V(T)$ is the variation of $T$ along the ray. The polar functions $v_k$ do not depend on $\bm x$, $\theta$, $\delta_R$, $\delta_N$ or the geometry, so $p_\eta$ bounds the separation rank uniformly in the Knudsen numbers.
\end{proposition}
\begin{proofof}
Since $0\leq\kappa_\tau\leq\mathrm{e}^{-\tau}$, one may take $u_k=0$ for $\tau\geq\tau_\eta=\ln\eta^{-1}$. Inserting the approximation of $\kappa_\tau$ into Eq.~\eqref{eq:kernel_rep} defines $X_k=\int_{[0,\tau_\eta)}u_k\dd m_{\bm x,\theta}$, with error at most $\|\kappa_0\|\bigl[\eta\,|m_{\bm x,\theta}|([0,\tau_\eta))+\int_{[\tau_\eta,\infty)}\mathrm{e}^{-\tau}\dd|m_{\bm x,\theta}|\bigr]$. The point masses and the density of $m_{\bm x,\theta}$ bound both terms by the right-hand side of Eq.~\eqref{eq:rank_error}. Without the cut-off at $\tau_\eta$, the part $\int_0^d|B|\dd t$ of the total variation of $m_{\bm x,\theta}$ would grow like $\delta_N$.
\end{proofof}

The proposition covers the exact transport solution for any given source, including the frozen-source problems of PGD--GSIS. For the DG solution $\phi$ still enters only through $\delta_C/\sin\phi$, and the same representation follows from
\[
 (\sin\phi\,G_\theta+\delta_C)^{-1}=\frac{1}{\sin\phi}\int_0^\infty\mathrm{e}^{-\delta_Ct/\sin\phi}\,\mathrm{e}^{-tG_\theta}\dd t,
\]
where the DG streaming matrix $G_\theta$ is accretive by Eq.~\eqref{eq:bh}.

The kernel rank is small in practice. A weighted SVD of $\kappa_\tau$ sampled at $\tau=0$ and $1201$ logarithmically spaced depths in $[10^{-8},10^3]$, with a maximum projection-error criterion, gives $p_\eta\approx5$, $10$ and $13$ for $\eta=10^{-2}$, $10^{-4}$ and $10^{-6}$ on the $40$-point Gauss--Legendre polar rule of Section~\ref{sec:results}. On a $160$-point rule the values are $5$, $13$ and $19$. These sampled ranks grow only moderately with the polar resolution, but they do not certify the continuous $p_\eta$.

\subsection{Rank in the transport regimes}\label{sec:regimes_rank}
Proposition~\ref{prop:rank} holds in every regime. How much of the bound $p_\eta$ a solution actually uses depends on how the measure $m_{\bm x,\theta}$ is distributed over optical depth, and this separates three cases.

\paragraph{Diffusive and hydrodynamic regimes} In the collision-dominated bulk, repeated integration by parts turns Eq.~\eqref{eq:characteristic} into a series in $\sin\phi/\delta_C$: the Chapman--Enskog expansion $I=E-\Kn_C\bm\Omega\cdot\nabla E+\mathcal O(\Kn_C^2)$ of the displaced equilibrium $E=(T+3\chi\bm\Omega\cdot\bm q)/(4\pi)$. Its terms are polynomial angular states, whose rank is fixed by their degree.
\begin{proposition}\label{prop:pn}
Let $f(\bm\Omega)$ be a polynomial of degree at most $N$ in the components of $\bm\Omega$ that is even in $\Omega_z=\cos\phi$. Then $f=\sum_{k=1}^{N+1}a_k(\theta)g_k(\phi)$, where $\{g_k\}$ is a basis of
\[
 G_N=\operatorname{span}\{\cos^{2j}\phi:\,2j\leq N\}\ \oplus\ \sin\phi\cdot\operatorname{span}\{\cos^{2j}\phi:\,2j\leq N-1\},
 \qquad \dim G_N=N+1 .
\]
Consequently a distribution whose angular dependence is, at every $\bm x$, such a polynomial has rank at most $N+1$ in the splitting~\eqref{eq:separated}, whereas its spherical-harmonic description has $(N+1)(N+2)/2$ coefficients per point.
\end{proposition}
\begin{proofof}
A monomial $\Omega_x^a\Omega_y^b\Omega_z^c$ with $c$ even equals $\cos^a\theta\sin^b\theta\cdot\sin^{n}\phi\cos^c\phi$, $n=a+b$. If $n$ is even, $\sin^n\phi=(1-\cos^2\phi)^{n/2}$ and the polar part is an even polynomial in $\cos\phi$ of degree $n+c\leq N$. If $n$ is odd, it is $\sin\phi$ times an even polynomial of degree $n-1+c\leq N-1$. Both lie in $G_N$, whose dimension is $\lfloor N/2\rfloor+\lfloor(N-1)/2\rfloor+2=N+1$. Expanding every polar part in a basis of $G_N$ and collecting the azimuthal coefficients gives the claim.
\end{proofof}

For $N=0$ this is the isotropic equilibrium, $T/(4\pi)=X\cdot1$ with $X=T/(4\pi)$. For $N=1$ it covers every $P_1$ state, in particular the displaced Callaway equilibrium and the Chapman--Enskog state of Fourier conduction:
\begin{equation}
 \frac{T}{4\pi}+\frac{3\bm\Omega\cdot\bm q}{4\pi}
 =\underbrace{\frac{T(\bm x)}{4\pi}}_{X_1}\cdot\underbrace{1}_{\Phi_1}
 +\underbrace{\bm s\cdot\bm q(\bm x)}_{X_2}\cdot\underbrace{\frac{3\sin\phi}{4\pi}}_{\Phi_2}.
 \label{eq:rank2}
\end{equation}
The azimuthal integrals $\int\bm s(\bm s\cdot\bm q)\dd\theta=\pi\bm q$ and $\mathsf M_2(\Phi_2)=1/\pi$ confirm that Eq.~\eqref{eq:rank2} returns $T$ and $\bm q$ through Eq.~\eqref{eq:macro_reconstruction}. The two collision-dominated regimes differ in the size of the flux. In the diffusive regime $\bm q=\mathcal O(\Kn_R)$, so to first order the state is the $P_1$ state~\eqref{eq:rank2}: rank two, and rank one where the flux vanishes. In the hydrodynamic regime $\bm q=\mathcal O(1)$, and the first streaming correction adds $-3\chi\Kn_C\sin^2\phi\,\bm s\cdot\nabla(\bm s\cdot\bm q)/(4\pi)$, which carries the viscous stress of the GK balance; the state is then $P_2$, of rank at most three. Each further order raises the degree, and with it the rank bound, by one. The conditional AP analysis of \ref{app:limits} uses these two polar spaces.

\paragraph{Ballistic regime} When $\delta_Cd\to0$ along rays of bounded in-plane length, the measure collapses to $\tau=0$ and $I$ tends to the boundary emission $I_b$, which does not depend on $\phi$: rank one, with $\Phi_1=1$. The approach is slow. The kernel differs from $\kappa_0$ mainly at grazing directions, $\sin\phi\lesssim\tau$, and $\|\kappa_\tau-\kappa_0\|/\|\kappa_0\|\approx(4\ln2/\pi)^{1/2}\tau^{1/2}$ for small $\tau$. This is consistent with the four to seven modes needed at the ballistic pair (Section~\ref{sec:rank}).

\paragraph{Transition regime and wall layers} At intermediate Knudsen numbers, and in the wall layers and corners of the collision-dominated regimes, the measure spreads over optical depths of order one. Neither expansion applies, and the rank is limited only by the kernel rank $p_\eta$ of Proposition~\ref{prop:rank}. Rank beyond two or three is therefore needed mainly there, and Section~\ref{sec:rank} confirms this ordering. A similar low-rank structure in the hydrodynamic and free-molecular limits was noted for the Shakhov model in~\cite{yin2026asymptotic}; Proposition~\ref{prop:rank} extends it to the transition regime with an explicit bound.

\section{PGD--GSIS construction and analysis}\label{sec:pgdgsis}
The factors are computed greedily: modes $1,\dots,m-1$ are retained and the pair $(X_m,\Phi_m)$ is determined by alternating Galerkin updates of the polar and the spatial factor. The Callaway gain term couples the unknown to its own macroscopic moments, and the two solvers compared in this paper treat this coupling differently. \emph{Pure PGD} is the Galerkin PGD of the complete linear equation~\eqref{eq:bte}: the gain term is evaluated from the kinetic moments of the current separated state, implicitly in the polar update and lagged by one sweep in the spatial update. \emph{PGD--GSIS} evaluates the gain term from a macroscopic source $\bm U_s=(T_s,\bm q_s)$ that is frozen during one polar/spatial pair and refreshed after the pair by the synthetic system~\eqref{eq:synthetic}. Section~\ref{sec:factor_eqs} derives the factor equations of both solvers, Section~\ref{sec:maps} gives the PGD--GSIS algorithm, Section~\ref{sec:cis_fourier} analyzes the convergence of both iterations, and Section~\ref{sec:ap} states a conditional AP property of PGD--GSIS.

\subsection{Factor equations of pure PGD and PGD--GSIS}\label{sec:factor_eqs}
Let $(a,b)_{x\theta}=\int_{\mathcal D}\int_0^{2\pi}ab\dd\theta\dd\bm x$ and let $b_h(V,U)$ be the upwind DG streaming form with interior upwind traces and outflow terms, the prescribed inflow being a separate load (\ref{app:dg}). Denote by $(T,\bm q)$ the macroscopic field that enters the gain term of Eq.~\eqref{eq:bte}. Testing Eq.~\eqref{eq:bte} with $X_m\Phi^*$ gives, pointwise in $\phi$,
\begin{equation}
 (\delta_C\alpha_{mm}+\beta_{mm}\sin\phi)\Phi_m
 =F_{T,m}+\sin\phi\,(F_{q,m}-\mu_m)
 -\sum_{r=1}^{m-1}(\delta_C\alpha_{mr}+\beta_{mr}\sin\phi)\Phi_r,
 \label{eq:polar}
\end{equation}
with $\alpha_{mr}=(X_m,X_r)_{x\theta}$, $\beta_{mr}=b_h(X_m,X_r)$, $F_{T,m}=\delta_C(X_m,T)_{x\theta}/(4\pi)$, $F_{q,m}=3\delta_N(X_m,\bm s\cdot\bm q)_{x\theta}/(4\pi)$, and $\mu_m$ the projection of the prescribed inflow on $X_m$. Testing with $X^*\Phi_m$ gives, for every azimuth,
\begin{equation}
 a_{mm}X_m+b_{mm}\bm s\cdot\nabla_hX_m
 =c_mT+d_m\bm s\cdot\bm q-\sum_{r=1}^{m-1}(a_{mr}X_r+b_{mr}\bm s\cdot\nabla_hX_r),
 \label{eq:spatial}
\end{equation}
\begin{equation}
 a_{mr}=\delta_C\!\int_0^\pi\!\Phi_m\Phi_r\dd\phi,\quad
 b_{mr}=\int_0^\pi\!\sin\phi\,\Phi_m\Phi_r\dd\phi,\quad
 c_m=\frac{\delta_C}{4\pi}\int_0^\pi\!\Phi_m\dd\phi,\quad
 d_m=\frac{3\delta_N}{4\pi}\mathsf T_m,
 \label{eq:wcoeff}
\end{equation}
where $\nabla_h$ denotes the DG streaming operator. On thermal inflow faces the projected condition $\sum_{r\leq m}b_{mr}X_r=\mathsf T_mT_b/(4\pi)$ enters through the upwind flux. Equation~\eqref{eq:spatial} is solved by DG sweeps, one per azimuth $\theta_j$, with local $n_p\times n_p$ systems on each triangle ($n_p=(p+1)(p+2)/2$). With elements ordered downstream, one sweep inverts the transport operator on the left except on periodic and adiabatic faces, whose incoming traces are lagged; sweeps are repeated until the relative change of $X_m$ falls below a tolerance or a cap $N_{\rm sw}$ is reached. The scaling of the pair is fixed by $X_m\leftarrow X_m/\|X_m\|$, $\Phi_m\leftarrow\|X_m\|\Phi_m$, applied together so that the product is unchanged. The two solvers differ only in the field $(T,\bm q)$ used in Eqs.~\eqref{eq:polar} and~\eqref{eq:spatial}.

\paragraph{Pure PGD} The gain term is evaluated from the kinetic moments~\eqref{eq:macro_reconstruction} of the current state, $(T,\bm q)=\mathcal K[I_m]=(T_m^{\kin},\bm q_m^{\kin})$, including the mode being computed. With $\alpha^T_{mr}=(X_m,Y_r)_{x\theta}/(4\pi)$ and $\alpha^q_{mr}=3(X_m,\bm s\cdot\bm Z_r)_{x\theta}/(4\pi)$, this gives $F_{T,m}=\delta_C\sum_{r\leq m}\alpha^T_{mr}\mathsf T_r$ and $F_{q,m}=\delta_N\sum_{r\leq m}\alpha^q_{mr}\mathsf Q_r$. The terms with $r=m$ depend on the unknown through $\mathsf T_m=\mathsf M_1(\Phi_m)$ and $\mathsf Q_m=\mathsf M_2(\Phi_m)$, and Eq.~\eqref{eq:polar} becomes
\begin{equation}
\begin{split}
 &(\delta_C\alpha_{mm}+\beta_{mm}\sin\phi)\Phi_m
 -\delta_C\alpha^T_{mm}\mathsf M_1(\Phi_m)-\delta_N\alpha^q_{mm}\sin\phi\,\mathsf M_2(\Phi_m)\\
 &\qquad=\sum_{r=1}^{m-1}\bigl[\delta_C\alpha^T_{mr}\mathsf T_r+\delta_N\alpha^q_{mr}\sin\phi\,\mathsf Q_r-(\delta_C\alpha_{mr}+\beta_{mr}\sin\phi)\Phi_r\bigr]-\mu_m\sin\phi .
\end{split}
 \label{eq:polar_implicit}
\end{equation}
The polar nodes are coupled through the moments $\mathsf M_1$ and $\mathsf M_2$: Eq.~\eqref{eq:polar_implicit} is an $N_\phi\times N_\phi$ system, diagonal plus rank two, which is solved directly. The macroscopic coupling of the current mode is therefore implicit in the polar update. In the spatial update, Eq.~\eqref{eq:spatial} with $(T,\bm q)=\mathcal K[I_m]$, the contributions $Y_m\mathsf T_m$ and $\bm Z_m\mathsf Q_m$ of the current mode are taken from the previous sweep, so each sweep is also one step of source iteration for the in-plane scattering problem of that mode. Pure PGD has no separate refresh step.

\paragraph{PGD--GSIS} The gain term is evaluated from the frozen source, $(T,\bm q)=\bm U_s$, which is refreshed once after each pair by the synthetic map (Section~\ref{sec:maps}). The $N_\phi$ polar nodes in Eq.~\eqref{eq:polar} then decouple and the polar update is a division. It is well defined: upwinding makes $b_h(X,X)\geq0$, so the coefficient on the left is at least $\delta_C\alpha_{mm}>0$. The right-hand side of Eq.~\eqref{eq:spatial} is fixed during the update; repeated sweeps only update the lagged periodic and adiabatic traces.

\begin{remark}[Spatial modes as in-plane transport solutions]\label{rem:opacity}
Equation~\eqref{eq:spatial}, divided by $b_{mm}$, is an in-plane transport equation. Its directions $\bm s(\theta)$ lie on the unit circle, and its effective extinction is
\begin{equation}
 \sigma_m=\frac{a_{mm}}{b_{mm}}=\delta_C\,\frac{\int_0^\pi\Phi_m^2\dd\phi}{\int_0^\pi\sin\phi\,\Phi_m^2\dd\phi}\ \geq\ \delta_C .
 \label{eq:opacity}
\end{equation}
A polar factor weighted toward grazing directions produces an optically thicker spatial mode. The greedy procedure thereby selects a short sequence of effective opacities, which is the discrete counterpart of approximating the kernel family $\kappa_\tau$ of Eq.~\eqref{eq:kernel_rep} by a few functions.
\end{remark}

\subsection{PGD--GSIS coupling and algorithm}\label{sec:maps}\label{sec:gsis}\label{sec:stopping}\label{sec:repcost}
PGD--GSIS couples each polar/spatial pair to a synthetic solve that refreshes $\bm U_s$ from the complete retained state $I_m$,
\begin{equation}
 \bm U_s\leftarrow\mathcal G[I_m]=(T^{\syn},\bm q^{\syn}),
 \label{eq:maps}
\end{equation}
where the synthetic map $\mathcal G$ solves the synthetic system~\eqref{eq:synthetic} with $\bm H$ evaluated from $I_m$. The kinetic map $\mathcal K$, the reconstruction~\eqref{eq:macro_reconstruction}, is used instead inside the factor updates of pure PGD; both solvers report $\mathcal K[I_m]$ as their kinetic moments. The synthetic map needs two ingredients from the separated state.

\paragraph{High-order moments from the PGD factors} The synthetic system needs $H_{ij}=\partial_kM_{ijk}$. Tested against $v_h$ on an element $K$,
\begin{equation}
 (v_h,H_{ij})_K
 =-\Bigl(\partial_kv_h,\int K_{ijk}I_m\dd\Omega\Bigr)_K
 +\Bigl\langle v_h,\int n_kK_{ijk}I_m^{\rm up}\dd\Omega\Bigr\rangle_{\partial K},
 \label{eq:weak_hot}
\end{equation}
where $I_m^{\rm up}$ is the upwind trace used by the kinetic sweep (interior neighbor, periodic partner, or wall emission). The face term is essential: dropping it would discard the jumps of the discontinuous kinetic state and change the discrete closure. For in-plane indices, $\int K_{ijk}X_r\Phi_r\dd\Omega=\int_0^{2\pi}X_r\bigl[\mathsf M_4(\Phi_r)s_is_js_k-\tfrac15\mathsf M_2(\Phi_r)(s_i\delta_{jk}+s_j\delta_{ik}+s_k\delta_{ij})\bigr]\dd\theta$, exactly under the discrete quadrature. Since the upwind side of a face depends only on $\theta$, the same holds for the face term. The load therefore costs $\mathcal O(mN_\theta N_x)$, with no $N_\phi$ factor, where $N_x=N_{\rm el}n_p$.

\paragraph{Synthetic solve and wall closure} Equation~\eqref{eq:synthetic} is written as a first-order system for $(T,\bm q)$ and the viscous flux and discretized by HDG~\cite{cockburn2009unified,nguyen2010stokes,nguyen2011navier} on the kinetic mesh with the same degree (\ref{app:hdg}). Element unknowns are eliminated locally and the global problem involves only the traces $(\hat T,\hat{\bm q})$ on the mesh skeleton; all factorizations depend on the mesh and the scattering rates only and are computed once. On physical walls the traces satisfy a linearized half-space law, $\widehat{\bm F}(\bm U)=\bm{\mathcal F}^{\kin}[I_m]+J_F(\bm U-\bm U^{\kin})$, in which $\bm{\mathcal F}^{\kin}$ is the kinetic wall flux of $I_m$ (energy flux and two momentum fluxes, requiring $\mathsf M_2$ and $\mathsf M_3$) and $J_F$ its derivative with respect to the Callaway equilibrium~\cite{Su2020JCP,liu2022fast}. The synthetic fields are the first three element components of the solution.

\paragraph{Algorithm and stopping criteria}
Algorithm~\ref{alg:pgd} has three nested loops, and none of its tests uses a reference solution. Spatial sweeps stop on the relative change of $X_m$ in the DG mass norm $\|\cdot\|_h$ (tolerance $\epsilon_{\rm sw}$, cap $N_{\rm sw}$). Alternating pairs stop when the relative change of the new mode's macroscopic amplitudes $\|Y_m\|_h|\mathsf T_m|$ and $\|Z_{m,i}\|_h|\mathsf Q_m|$ falls below $\epsilon_{\rm ALS}$, or after $L_{\max}$ pairs. Enrichment stops when
\begin{equation}
 r_{\rm en}^{(m)}=\frac{\|\bm W_m\|_h}{\bigl\|\sum_{r\leq m}\bm W_r\bigr\|_h}<\epsilon_{\rm en},\qquad \bm W_r=(Y_r\mathsf T_r,\bm Z_r\mathsf Q_r),
 \label{eq:enrichment}
\end{equation}
for some $m>1$, or at $M_{\max}$ modes. All three are increment tests. Section~\ref{sec:cis_fourier} shows their limitations for pure PGD. Under the synthetic map they work well in the reported experiments, but a small modal amplitude is not an error bound, because the modes are not orthogonal. In particular, these tests do not certify the projected residual and source-matching conditions of Section~\ref{sec:ap}; those conditions are stated separately from the stopping rules used for the runs. Pure PGD uses the same loops, tests and caps with three changes: no HDG data are assembled, the polar update is Eq.~\eqref{eq:polar_implicit}, and the spatial sweeps lag the current-mode moments; there is no refresh step.

\begin{algorithm}[tb]
\caption{Fixed-parameter PGD--GSIS}
\label{alg:pgd}
\begin{algorithmic}[1]
\REQUIRE mesh, angular quadrature, boundary data, $(\delta_R,\delta_N)$, tolerances and caps
\STATE Assemble DG data; factorize the local and global HDG systems and the wall Jacobians.
\STATE Set $\bm U_s=(T_s,\bm q_s)=0$.
\FOR{$m=1,\ldots,M_{\max}$}
 \STATE Initialize $X_m$.
 \REPEAT
  \STATE Polar update~\eqref{eq:polar} with $\bm U_s$ frozen; compute $\mathsf M_k(\Phi_m)$, $k=1,\dots,4$, and the coefficients~\eqref{eq:wcoeff}.
  \STATE Spatial update~\eqref{eq:spatial} with $\bm U_s$ frozen: DG sweeps until the sweep test holds.
  \STATE Normalize the pair.
  \STATE Synthetic refresh from the complete retained state: $\bm U_s\leftarrow\mathcal G[I_m]$.
 \UNTIL{the amplitude test holds or $L_{\max}$ pairs have been performed}
 \STATE Retain $(X_m,\Phi_m)$; exit if $r_{\rm en}^{(m)}<\epsilon_{\rm en}$ and $m>1$.
\ENDFOR
\ENSURE factors $\{X_r,\Phi_r\}$, kinetic moments $\mathcal K[I_m]$ and synthetic fields $\mathcal G[I_m]$
\end{algorithmic}
\end{algorithm}

\paragraph{Storage and computational cost}
For a single parameter the distribution needs $N_xN_\theta N_\phi$ values in full order and $m(N_xN_\theta+N_\phi)$ in separated form. With $N_x=2000$, $N_\theta=80$, $N_\phi=40$ and $m=30$ these are $6.4\times10^6$ and $4.8\times10^6$: the representation itself saves little at this angular resolution. What changes is the scaling. The separated storage is essentially independent of $N_\phi$, and, in the parametric setting of Section~\ref{sec:parametric} with $N_\delta$ parameter nodes, it becomes $m(N_xN_\theta+N_\phi N_\delta)$ against $N_xN_\theta N_\phi N_\delta$; at $N_\delta=441$ and the same rank, $5.3\times10^6$ against $2.8\times10^9$. Peak solver memory also includes DG projections, the HDG factorizations and the synthetic fields; these do not depend on $N_\phi$ but do scale with $N_\delta$ (Section~\ref{sec:param_cost}).

Per alternating pair, the spatial update costs $\mathcal O(N_\theta N_{\rm el}n_p^3)$ for the local factorizations plus $\mathcal O(N_\theta N_{\rm el}n_p^2)$ per sweep and $\mathcal O(mN_\theta N_x)$ for the retained-mode terms, with no $N_\phi$ factor; the polar update costs $\mathcal O(mN_\phi)$, plus one dense $N_\phi\times N_\phi$ solve in pure PGD; the synthetic refresh costs one high-order load, $\mathcal O(mN_\theta N_x)$ in the moment form above, and one backsolve. A full-order sweep costs $\mathcal O(N_\theta N_\phi N_{\rm el}n_p^2)$. At equal numbers of source updates, PGD--GSIS therefore gains over full-order GSIS only when $N_\phi$ exceeds the rank by a sufficient factor; with $N_\phi=40$ and ranks $3$--$14$ the two are comparable (Section~\ref{sec:cost}).

\subsection{Convergence and kinetic--synthetic consistency}\label{sec:cis_fourier}\label{sec:consistency}
Consider a Fourier error mode $\mathrm{e}^{ikx}$ in a periodic homogeneous medium (\ref{app:fourier}). Define $\mathcal S\bm U=(\delta_CT+3\delta_N\bm\Omega\cdot\bm q)/(4\pi)$ and $\mathcal T=\bm\Omega\cdot\nabla+\delta_C$. The full-order operators $A_{\mathcal M}=\mathcal M\circ\mathcal T^{-1}\circ\mathcal S$, $\mathcal M\in\{\mathcal K,\mathcal G\}$, are the error-propagation operators of source iteration and of GSIS~\cite{liu2022fast}; they serve as references for the two PGD iterations.

\paragraph{Pure PGD} The implicit polar update~\eqref{eq:polar_implicit} balances the current mode against its own moments only as seen through its fixed spatial shape $X_m$. The streaming of that mode enters only through $\beta_{mm}=b_h(X_m,X_m)$, which consists of outflow and interface-jump terms (\ref{app:dg}) and vanishes for a smooth mode in a periodic bulk. The polar update therefore adjusts the angular shape and the amplitude of the mode, but it does not move macroscopic information across the domain. Only the spatial update does, and there the moments of the current mode are lagged. For a fixed polar factor $\Phi$, Eq.~\eqref{eq:spatial} is an in-plane transport problem with extinction $a/b$ (Remark~\ref{rem:opacity}) and in-scattering $c\mathsf T\,Y+d\mathsf Q\,\bm s\cdot\bm Z$, and each sweep is one step of source iteration for it. For the error mode $\mathrm{e}^{ikx}$, a sweep acts linearly on the azimuthal moments $(\hat Y,\hat{\bm Z})$ of the error; we denote its spectral radius by $\rho_{\rm sw}$ (\ref{app:fourier}). Table~\ref{tab:fourier} lists $1-\rho_{\rm sw}$ for the two polar factors of the equilibrium~\eqref{eq:rank2}: $\Phi=1$, which carries the temperature, and $\Phi=\sin\phi$, which carries the heat flux. These margins are of the same order as the margin $1-\rho(A_{\mathcal K})$ of full-order source iteration, whose leading estimates are $\Kn_C\Kn_Rk^2/3$ for a Fourier temperature mode and $\Kn_C(\delta_R+\Kn_Ck^2/5)$ for a transverse hydrodynamic flux mode~\cite{liu2022fast,adams2002fast}. The temperature factor is the slower one at the diffusive pairs; the flux factor is the slower one at the hydrodynamic pairs, where normal scattering nearly conserves momentum.

Reducing a slow error component by one decade therefore takes $\ln10/(1-\rho_{\rm sw})$ sweeps: about $3\times10^4$ and $10^4$ at the two diffusive pairs and $8\times10^3$ at $(100,0.01)$, against a budget of $N_{\rm sw}L_{\max}=50$ sweeps per mode in Section~\ref{sec:results}. The increment tests do not detect this. Once the error is dominated by its slow component, one sweep changes the state by only $1-\rho_{\rm sw}$ times that error, so the sweep test with $\epsilon_{\rm sw}=10^{-3}$ is passed whenever $1-\rho_{\rm sw}<10^{-3}$, although the error has hardly been reduced. Enrichment then fits new modes to a state whose macroscopic part has not converged, and errors in the $\ell\leq1$ content of the state re-enter the gain term through $\mathcal K$ with unit gain (Proposition~\ref{prop:fourier}(i) below). The slow components are close to $P_1$ states of rank at most two (Proposition~\ref{prop:pn}), so low-rank representability and slow convergence are compatible. At the ballistic pair, by contrast, $1-\rho_{\rm sw}\approx0.99$.

\paragraph{PGD--GSIS} Let $\bm U^n$ be the source error before alternating pair $n$, $I_{\rm ex}^n=\mathcal T^{-1}\mathcal S\bm U^n$ the exact frozen-source response, and $e^n=I_m^n-I_{\rm ex}^n$ the representation error, which combines truncation, incomplete alternation and stale earlier modes. Executing Algorithm~\ref{alg:pgd} in this setting (\ref{app:fourier}) gives
\begin{equation}
 \bm U^{n+1}=A_{\mathcal G}\bm U^n+\mathcal G[e^n].
 \label{eq:error_split}
\end{equation}
The same identity with $\mathcal K$ in place of $\mathcal G$ holds for a frozen-source alternation refreshed by kinetic moments, whose leading operator is then full-order source iteration (Section~\ref{sec:param_cost}). The effect of PGD errors on PGD--GSIS is therefore determined by their image under the synthetic map.

\begin{proposition}\label{prop:fourier}
(i) $\mathcal K[e]=(\int e\dd\Omega,\int\bm\Omega e\dd\Omega)$ reads the angular content of $e$ of degree $\ell\leq1$, with unit gain.
(ii) In a periodic domain, $\mathcal G[e]$ depends on $e$ only through $M_{ijk}[e]=\int K_{ijk}e\dd\Omega$. Since $K_{ijk}$ is the symmetric trace-free part of $\bm\Omega\otimes\bm\Omega\otimes\bm\Omega$, a spherical harmonic of degree three, only the $\ell=3$ content of $e$ enters. For the mode $\mathrm{e}^{ikx}$,
\begin{equation}
 \mathcal G[e]=\Bigl(3ik\Kn_C\,M_{xxx}[e],\ 0,\ -\frac{\Kn_Ck^2}{\delta_R+\Kn_Ck^2/5}M_{yxx}[e]\Bigr).
 \label{eq:G_fourier}
\end{equation}
\end{proposition}
\begin{proofof}
Part (i) is the definition~\eqref{eq:macro_reconstruction}. For (ii), $\mathcal G$ is linear and, in a periodic domain, its only kinetic input is $\bm H=\nabla\cdot\bm M$. Substituting $\nabla\to ik\bm e_x$ into Eq.~\eqref{eq:synthetic} gives $\hat q_x=0$ from $\nabla\cdot\bm q=0$ and Eq.~\eqref{eq:G_fourier} from the two momentum equations.
\end{proofof}

Since $\Kn_Ck^2/(\delta_R+\Kn_Ck^2/5)=5(\ell_\GK k)^2/[1+(\ell_\GK k)^2]$, the gain on the $\ell=3$ content is $3k\Kn_C$ for the temperature and at most $5$ for the transverse flux. Both gains are small only in the Fourier regime, $k\Kn_C\ll1$ and $\ell_\GK k\ll1$; in the hydrodynamic regime the flux gain is of order one, and only the restriction to $\ell=3$ remains. The degree-three filtering in part (ii) concerns the periodic bulk. The wall law of Section~\ref{sec:gsis} adds half-range moment inputs with no explicit $\Kn_C$ prefactor. Defining $\gamma_{\mathcal G}^n=\|\mathcal G[e^n]\|/\|\bm U^n\|$, Eq.~\eqref{eq:error_split} gives $\|\bm U^{n+1}\|\leq(\|A_{\mathcal G}\|+\gamma_{\mathcal G}^n)\|\bm U^n\|$. This bound uses the operator norm; the spectral radii and median perturbations in Table~\ref{tab:fourier} describe the observed iteration but do not establish a uniform contraction bound.

\begin{table}[tb]
\centering
\caption{Fourier analysis for an error mode $\mathrm{e}^{i\pi x}$ ($N_\theta\times N_\phi=80\times40$). $1-\rho(A_{\mathcal K})$: margin of full-order source iteration; $1-\rho_{\rm sw}^{T}$, $1-\rho_{\rm sw}^{q}$: margin of one lagged sweep of the pure-PGD spatial update for the polar factors $\Phi=1$ and $\Phi=\sin\phi$; $\rho(A_{\mathcal G})$: spectral radius of full-order GSIS; $\|e\|$: median relative error $\|I_m^n-I_{\rm ex}^n\|/\|I^n_{\rm ex}\|$ of the separated state in Algorithm~\ref{alg:pgd} ($M_{\max}=20$, $L_{\max}=10$); $\gamma_{\mathcal G}$: median over pairs of $\|\mathcal G[e^n]\|/\|\bm U^n\|$; $n_{12}$: pairs needed to reduce the source error by $10^{12}$.}
\label{tab:fourier}
\footnotesize
\setlength{\tabcolsep}{4pt}
\begin{tabular}{l r rr rrrr}
\toprule
 & full order & \multicolumn{2}{c}{pure PGD, spatial sweep} & \multicolumn{4}{c}{PGD--GSIS}\\
\cmidrule(lr){2-2}\cmidrule(lr){3-4}\cmidrule(lr){5-8}
$(\Kn_R,\Kn_N)$ & $1-\rho(A_{\mathcal K})$ & $1-\rho_{\rm sw}^{T}$ & $1-\rho_{\rm sw}^{q}$ & $\rho(A_{\mathcal G})$ & $\|e\|$ & $\gamma_{\mathcal G}$ & $n_{12}$\\
\midrule
$(0.01,0.01)$ & $1.6\times10^{-4}$ & $8.0\times10^{-5}$ & $1.8\times10^{-4}$ & $4.2\times10^{-9}$ & $3.2\times10^{-3}$ & $6.9\times10^{-6}$ & $3$\\
$(0.01,100)$  & $3.3\times10^{-4}$ & $2.0\times10^{-4}$ & $3.6\times10^{-4}$ & $6.7\times10^{-8}$ & $6.0\times10^{-3}$ & $2.8\times10^{-5}$ & $3$\\
$(0.1,10)$    & $3.1\times10^{-2}$ & $1.9\times10^{-2}$ & $3.3\times10^{-2}$ & $5.8\times10^{-4}$ & $5.8\times10^{-2}$ & $2.2\times10^{-3}$ & $5$\\
$(10,0.1)$    & $2.8\times10^{-2}$ & $9.3\times10^{-2}$ & $2.7\times10^{-2}$ & $1.4\times10^{-2}$ & $2.2\times10^{-2}$ & $2.8\times10^{-3}$ & $7$\\
$(100,0.01)$  & $3.0\times10^{-4}$ & $8.0\times10^{-4}$ & $2.8\times10^{-4}$ & $1.5\times10^{-4}$ & $2.2\times10^{-3}$ & $3.2\times10^{-5}$ & $4$\\
$(100,100)$   & $0.99$ & $0.99$ & $0.99$ & $0.79$ & $0.59$ & $3.3\times10^{-2}$ & $133$\\
\bottomrule
\end{tabular}
\end{table}

\paragraph{Effect of the synthetic refresh} At the collision-dominated pairs, Table~\ref{tab:fourier} gives $\rho(A_{\mathcal G})\leq1.4\times10^{-2}$ and a $10^{12}$ source-error reduction in $3$--$7$ pairs, but the reason differs between regimes. At the three diffusive pairs the synthetic image of the representation error is much smaller than its kinetic image: the median of $\|\mathcal G[e^n]\|/\|\mathcal K[e^n]\|$ is $5.0\times10^{-3}$, $1.1\times10^{-2}$ and $8.7\times10^{-2}$, as the small gains of Proposition~\ref{prop:fourier} predict. At the two hydrodynamic pairs $\ell_\GK k\approx1.4$, the flux gain is about $3.3$, and the two images are comparable (median ratio $1.6$). There the fast convergence comes from the small full-order factor $\rho(A_{\mathcal G})$ together with the small PGD error, $\gamma_{\mathcal G}\leq2.8\times10^{-3}$, not from filtering. At the ballistic pair the roles reverse: $\rho(A_{\mathcal G})$ approaches its limit $4/5$ (\ref{app:fourier}) and PGD--GSIS needs $133$ pairs, whereas the margins of source iteration and of the pure-PGD sweep are close to one. This small GSIS margin at large $\Kn_C$ is one reason why the parametric construction needs the joint polar update of Section~\ref{sec:param_cost}.

\paragraph{Consistency of kinetic and synthetic fields}
At a self-consistent solution of the continuous kinetic problem with consistent wall data, $\mathcal G[I_m]=\mathcal K[I_m]$. For an approximate state, define the moment residuals of Eq.~\eqref{eq:moment_system} by $r_0=\nabla\cdot\bm q^{\kin}$, $\bm r_1=\tfrac13\nabla T^{\kin}+\nabla\cdot\bm N^{\kin}+\delta_R\bm q^{\kin}$ and $\bm r_2=\tfrac25\partial_{\langle j}q^{\kin}_{i\rangle}+\nabla\cdot\bm M^{\kin}+\delta_C\bm N^{\kin}$. Eliminating $\bm N^{\kin}$ shows that the defect $\mathcal G[I_m]-\mathcal K[I_m]$ solves the synthetic system with sources $-r_0$ and $-\bm r_1+\Kn_C\nabla\cdot\bm r_2$, plus a boundary contribution. For a nonzero periodic Fourier mode with $r_0=0$, $\Delta T=-3r_{1x}/(ik)+3\Kn_Cr_{2xx}$ exhibits amplification of second-moment residuals at large $\Kn_C$. Consistency therefore requires residual control; enrichment alone does not certify it. Section~\ref{sec:ap} quantifies bulk matching at small $\Kn_C$, while Section~\ref{sec:param_error} discusses node-dependent residuals in a shared parametric basis.

\subsection{Conditional asymptotic-preserving property}\label{sec:ap}
\ref{app:limits} first expands the frozen-source pair formally in $\Kn_C$, as~\cite{yin2026asymptotic} did for the Shakhov model. To leading order the pair is an alternating least-squares iteration for the singular value decomposition of the local equilibrium, with a contraction factor independent of $\Kn_C$, and the first-order Chapman--Enskog state has rank three. The expansion assumes that the rows of the frozen earlier modes hold, which the greedy construction does not ensure: in a Fourier test, three modes span the required polar space but leave an $\mathcal O(\Kn_C)$ error, which only a fourth mode removes. The appendix therefore defines three indicators: the complete projected transport residual $\eta_n$, the polar-space approximation error $\zeta_n$ for the equilibrium and its first streaming correction, and the macroscopic source increment $\omega_n$ (Eq.~\eqref{eq:ap_indicators}). Under uniform bulk regularity, Proposition~\ref{prop:ap} derives
\begin{equation}
 \|\mathcal K[I_m^n]-\mathcal G[I_m^n]\|_2
 \leq C\bigl(\Kn_C^2+\eta_n+\zeta_n+\omega_n\bigr).
 \label{eq:ap_main_matching}
\end{equation}
Thus $\eta_n+\zeta_n+\omega_n=\mathcal O(\Kn_C^{1+\sigma_{\rm AP}})$, $0<\sigma_{\rm AP}\leq1$, is sufficient to recover the leading Fourier flux, including its coefficient $\Kn_R/3$, in the diffusive limit (Corollary~\ref{cor:ap_diffusive}), and the viscous GK balance, or a Brinkman/Stokes problem in a distinguished scaling, in the hydrodynamic limit (Corollary~\ref{cor:ap_hydro}). The matching estimate is derived from the projected spatial equations and the preceding synthetic balance. In the ballistic limit the gain term carries the factor $\delta_C$, so the synthetic source, whose response to representation errors grows like $\Kn_C$, reaches the kinetic moments with an $\mathcal O(1)$ gain, and the frozen-source pair reproduces free streaming with one polar mode (Proposition~\ref{prop:ballistic}).

The collision-dominated results are conditional bulk statements. The greedy spatial update enforces only its current-mode row; frozen rows and polar-space completeness must also be controlled. The amplitude tests of Algorithm~\ref{alg:pgd} do not imply these bounds, and the reported fixed tolerances and caps are not claimed to certify AP for arbitrarily small Knudsen numbers. Parametric certification requires the bounds node by node. Wall layers, compatibility of the discrete DG/HDG moments, angular quadrature and a uniform iteration budget require further analysis. Preservation of the limiting constitutive law is distinct from the acceleration measured in Section~\ref{sec:cis_fourier}.

\section{Parametric PGD--GSIS over the Knudsen numbers}\label{sec:parametric}
\subsection{Representation and factor equations}\label{sec:param_cost}
Let $\boldsymbol\delta=(\delta_R,\delta_N)$ range over a parameter set $\mathcal P$ with a positive measure $\dd\nu$. The parametric representation attaches the parameters to the polar factor,
\begin{equation}
 I_m(\bm x,\theta,\phi;\boldsymbol\delta)=\sum_{r=1}^mX_r(\bm x,\theta)\,\Phi_r(\phi,\boldsymbol\delta),
 \qquad
 T_m^{\kin}=\sum_{r=1}^mY_r(\bm x)\mathsf T_r(\boldsymbol\delta),\quad
 \bm q_m^{\kin}=\sum_{r=1}^m\bm Z_r(\bm x)\mathsf Q_r(\boldsymbol\delta).
 \label{eq:parameterized}
\end{equation}
The spatial factors are shared by all parameter values; $(\phi,\delta_R,\delta_N)$ form a single joint coordinate, which is natural because $\delta_C$ and $1/\sin\phi$ enter the kernel of Eq.~\eqref{eq:kernel_rep} as a product. Each inverse Knudsen number takes $N$ logarithmically spaced nodes on $[\delta_{\min},\delta_{\max}]$, and $\dd\nu$ is the tensor product of the normalized measures $\dd(\log\delta)/\log(\delta_{\max}/\delta_{\min})$, discretized by the trapezoidal rule in $\log\delta$. The constructions below use $\Kn_R,\Kn_N\in[10^{-2},10^2]$ with $N=21$ (0.2 decades, $N_\delta=441$ nodes). The measure is part of the method: it decides how much each region of $\mathcal P$ influences the shared basis.

With the source frozen at every node, the polar update is Eq.~\eqref{eq:polar} evaluated node by node: the nodes are independent and each update is again a division. The spatial update is Eq.~\eqref{eq:spatial} with parameter-averaged coefficients,
\begin{equation}
\begin{gathered}
 \bar a_{mr}=\int_{\mathcal P}\!\int_0^\pi\delta_C\Phi_m\Phi_r\dd\phi\dd\nu,\qquad
 \bar b_{mr}=\int_{\mathcal P}\!\int_0^\pi\sin\phi\,\Phi_m\Phi_r\dd\phi\dd\nu,\\
 \bar a_{mm}X_m+\bar b_{mm}\bm s\cdot\nabla_hX_m
 =S_{T,m}+\bm s\cdot\bm S_{q,m}-\sum_{r=1}^{m-1}(\bar a_{mr}X_r+\bar b_{mr}\bm s\cdot\nabla_hX_r),
\end{gathered}
 \label{eq:param_spatial}
\end{equation}
with $S_{T,m}=\frac{1}{4\pi}\int_{\mathcal P}\int_0^\pi\delta_C\Phi_mT_s\dd\phi\dd\nu$ and $\bm S_{q,m}=\frac{3}{4\pi}\int_{\mathcal P}\int_0^\pi\delta_N\sin\phi\,\Phi_m\bm q_s\dd\phi\dd\nu$; boundary loads are projected likewise. By Remark~\ref{rem:opacity}, each shared spatial mode is a two-dimensional transport solution at a \emph{single} effective extinction $\bar a_{mm}/\bar b_{mm}$, the ratio of the $\Phi_m^2$-weighted averages of $\delta_C$ and $\sin\phi$ over the polar angle and the parameter box. Covering four decades of $\delta_C$ therefore requires modes spread over the corresponding range of effective opacities. The synthetic refresh is performed independently at every node with that node's HDG operator.

The separated distribution is small (Section~\ref{sec:repcost}), but the synthetic operator is stored per node: the local HDG factors alone take $(7n_p)^2N_{\rm el}$ values per node, $7.84$\,MB for $p=3$ and $N_{\rm el}=200$, and the trace influence matrices and global factorizations add to this. For $N_\delta=441$ these per-node operators dominate the peak memory (Section~\ref{sec:param_results}). They are independent of the rank and could be computed on the fly or shared between neighboring nodes; we have not pursued this.

\paragraph{Joint polar update} With the spatial factors fixed, all retained polar factors are re-solved against the refreshed source:
\begin{equation}
 \sum_{r=1}^m(\delta_C\alpha_{kr}+\beta_{kr}\sin\phi)\Phi_r
 =F_{T,k}+\sin\phi\,(F_{q,k}-\mu_k),\qquad k=1,\ldots,m,
 \label{eq:joint_polar}
\end{equation}
an $m\times m$ system at every $(\phi,\boldsymbol\delta)$. At each parameter node it is the Galerkin reduced model of the frozen-source transport problem on the current spatial basis, so the nodal state becomes the Galerkin approximation in $\operatorname{span}\{X_1,\dots,X_m\}\otimes L^2(0,\pi)$. Old factors change, the construction is no longer purely greedy, and the amplitude test is evaluated on the complete retained state.

The joint update is required for accuracy over the whole parameter box. Without it, $\Phi_r(\cdot,\boldsymbol\delta)$ keeps the source that was available when mode $r$ was computed. When $\bm U_s(\boldsymbol\delta)$ changes, the state can follow only through the current mode, whose spatial factor is shared by all nodes and fitted to their parameter average~\eqref{eq:param_spatial}. At nodes that this factor fits poorly, the change of the frozen-source response becomes representation error that does not decrease as modes are added. In the Fourier regime the synthetic map damps such errors (Proposition~\ref{prop:fourier}), but at large $\Kn_C$ the kinetic--synthetic consistency gain is $\mathcal O(\Kn_C)$ (Section~\ref{sec:consistency}) and the contraction margin of full-order GSIS shrinks to $1-\rho(A_{\mathcal G})\to1/5$ (Table~\ref{tab:fourier}). The perturbation $\gamma_{\mathcal G}^n$ in Eq.~\eqref{eq:error_split} can then exceed that margin: the construction loses accuracy at large Knudsen numbers and can diverge at the rarefied corner (Section~\ref{sec:param_results}). With the joint update, the nodal state is the Galerkin solution on the current spatial basis, so the representation error is governed by that basis rather than by stale polar factors. It is not guaranteed to decrease monotonically, because a Galerkin projection of a transport operator is not a best approximation.

Other remedies are possible. The parameter domain can be divided into subregions, for example by ranges of $\Kn_C$, each with its own spatial basis; or the polar factors can be re-projected after the construction, or the modes recompressed by a singular value decomposition of the nodal solutions. Such a posteriori steps can remove stale polar coefficients, but they cannot repair a spatial basis whose modes were fitted to the residuals of stale nodal states. The in-loop update improves both, because each new spatial mode is fitted to the residual of nodal Galerkin solutions on the current basis. We therefore use the joint update; it also keeps a single construction, and its algebraic systems are small. Its matrices need only $\alpha_{kr}$ and $\beta_{kr}$, which do not depend on $\boldsymbol\delta$, and its loads need the projections of the refreshed source on the $m$ spatial factors, $\mathcal O(mN_xN_\delta)$ operations. The $m\times m$ solves then cost $\mathcal O(m^3N_\phi N_\delta)$ operations, small compared with the $N_\delta$ synthetic backsolves and the parameter-averaged sweeps of the same pair (Section~\ref{sec:param_results}).

\paragraph{Parametric pure PGD} For comparison, a parametric pure PGD can use the joint update in two forms. With a frozen kinetic source, $\bm U_s\leftarrow\mathcal K[I_m]$ once per pair, Eq.~\eqref{eq:joint_polar} applies unchanged, and Eq.~\eqref{eq:error_split} with $\mathcal K$ in place of $\mathcal G$ shows that the outer iteration is then full-order source iteration, whose margin vanishes at the collision-dominated nodes (Table~\ref{tab:fourier}). With the implicit coupling of Section~\ref{sec:factor_eqs}, the joint update reads, at every parameter node,
\begin{equation}
\begin{split}
 \sum_{r=1}^m\Bigl[(\delta_C\alpha_{kr}+\beta_{kr}\sin\phi)\Phi_r
 -\delta_C\alpha^T_{kr}\mathsf M_1(\Phi_r)-\delta_N\alpha^q_{kr}\sin\phi\,\mathsf M_2(\Phi_r)\Bigr]
 =-\mu_k\sin\phi,\qquad k=1,\ldots,m.
\end{split}
 \label{eq:joint_polar_implicit}
\end{equation}
It couples all polar nodes through $\mathsf M_1$ and $\mathsf M_2$: an $mN_\phi\times mN_\phi$ system, block diagonal in $\phi$ apart from a coupling of rank $2m$, which is the Galerkin reduced model of the full kinetic problem on the current spatial basis. No source iteration then remains in the polar factors, but the spatial factors are still built by the lagged sweeps of Section~\ref{sec:cis_fourier}, which move macroscopic information by about one mean free path per sweep. In our tests, parametric pure PGD with the frozen kinetic source did not reach the correct solution with $30$ modes; with the implicit joint update it did, but only with considerably more modes and alternating pairs than PGD--GSIS, whose macroscopic source comes from the synthetic equations and whose joint update needs only $m\times m$ solves. Algorithm~\ref{alg:parametric} summarizes the PGD--GSIS construction.

\begin{algorithm}[tb]
\caption{Parametric PGD--GSIS}
\label{alg:parametric}
\begin{algorithmic}[1]
\REQUIRE parameter nodes and weights, mesh and quadrature, tolerances and caps
\STATE Factorize the local and global HDG systems and wall Jacobians at every parameter node.
\FOR{$m=1,\ldots,M_{\max}$}
 \STATE Initialize $X_m$.
 \REPEAT
  \STATE Polar update~\eqref{eq:polar} at every node, source frozen.
  \STATE Spatial update~\eqref{eq:param_spatial}, source frozen; normalize the pair.
  \STATE Synthetic refresh $\bm U_s(\boldsymbol\delta)\leftarrow\mathcal G_{\boldsymbol\delta}[I_m]$ at every node.
  \STATE Joint polar update~\eqref{eq:joint_polar} of all retained polar factors at every node.
 \UNTIL{the amplitude test holds or $L_{\max}$ pairs have been performed}
 \STATE Retain $(X_m,\Phi_m)$; exit if $r_{\rm en}^{(m)}<\epsilon_{\rm en}$.
\ENDFOR
\ENSURE shared spatial factors $X_r$, nodal polar factors $\Phi_r(\cdot,\boldsymbol\delta_j)$, synthetic fields
\end{algorithmic}
\end{algorithm}

\subsection{Parametric rank and error sensitivity}\label{sec:param_error}
Equation~\eqref{eq:parameterized} places every $T_m^{\kin}(\cdot,\boldsymbol\delta)$ in the same space $\operatorname{span}\{Y_1,\dots,Y_m\}$. For the temperature family $\mathcal M_T=\{T(\cdot;\boldsymbol\delta)\}_{\boldsymbol\delta\in\mathcal P}$, uniform accuracy $\varepsilon_T$ therefore requires $d_m(\mathcal M_T)\leq\varepsilon_T$. The Kolmogorov width $d_m$ is the smallest worst-case approximation error over all $m$-dimensional spatial spaces in the chosen norm. An analogous condition holds for the flux. The common basis must capture changes from harmonic fields to nearly isothermal bulks with wall jumps, which can require more modes than the individual fixed-parameter problems.

For Fourier states with $\|\bm q\|\asymp\Kn_R$, a distribution error $e=I_m-I$ can produce a relative kinetic-flux error of order $\|e\|/\Kn_R$. The parameter-averaged Galerkin equations do not directly control these relative errors at individual nodes. This motivates nodewise error checks and, if needed, a parameter measure that emphasizes the small-$\Kn_R$ edge. The bulk matching conditions of Proposition~\ref{prop:ap} must likewise hold node by node; a small parameter-weighted residual alone does not establish them.

At large $\Kn_C$, second-moment residuals can enter the synthetic temperature with gain $\mathcal O(\Kn_C)$ (Section~\ref{sec:consistency}), making the ballistic corner another sensitive region. The joint update of Section~\ref{sec:param_cost} removes the stale-factor part of these residuals; with it, all our constructions remained stable over the whole parameter box (Section~\ref{sec:param_results}).

Nodal macroscopic reconstruction costs $\mathcal O(mN_x)$. Between nodes, interpolation of $\mathsf T_r$ and $\mathsf Q_r$ adds an error separate from the nodal approximation; Section~\ref{sec:param_results} assesses it independently by withholding nodes.

\section{Numerical examples}\label{sec:results}
\subsection{Numerical setup and error measures}
The fixed-parameter tests use six pairs $(\Kn_R,\Kn_N)$: $(0.01,0.01)$, $(0.01,100)$, $(0.1,10)$, $(10,0.1)$, $(100,0.01)$ and $(100,100)$. In the present benchmarks, the first three are predominantly diffusive, the next two probe hydrodynamic transport, and the last is ballistic. The first five have $\Kn_C\leq0.1$, whereas the ballistic pair has $\Kn_C=50$. For the two hydrodynamic pairs, Eq.~\eqref{eq:gk} gives $\ell_\GK\approx0.44$ and $0.45$, both larger than the channel height $H=0.2$, so viscous effects are important in Poiseuille flow. These regime labels describe representative behavior; geometry and boundary conditions determine how it appears in each problem.

The spatial factor uses polynomials of degree $p=3$ on unstructured triangles: $200$ elements for the cavity and $80$ for the channels. The azimuth uses $N_\theta=80$ nodes (Gauss--Legendre rules on $[0,\pi]$ and $[\pi,2\pi]$) and the polar angle $N_\phi=40$ Gauss--Legendre nodes on $[0,\pi]$. The fixed-parameter runs use at most $N_{\rm sw}=5$ sweeps per spatial update, $L_{\max}=10$ alternating pairs per mode and $M_{\max}=20$ modes, with tolerances $\epsilon_{\rm sw}=\epsilon_{\rm ALS}=10^{-3}$. Figures~\ref{fig:comparison}--\ref{fig:amplitude_decay} retain all $20$ modes; Section~\ref{sec:cost} stops enrichment at $\epsilon_{\rm en}=10^{-3}$.
Reference solutions are computed by full-order GSIS with the same spatial and angular discretization; they enter only the reported errors,
\begin{equation}
 E_T=\frac{\|T_m^{\kin}-T_{\FR}\|_h}{\max(\|T_{\FR}\|_h,s_T)},\qquad
 E_q=\frac{\|\bm q_m^{\kin}-\bm q_{\FR}\|_h}{\max(\|\bm q_{\FR}\|_h,s_q)},
 \label{eq:reference_errors}
\end{equation}
with small positive floors $s_T,s_q$. Error maps show $|T-T_{\FR}|/\max|T_{\FR}|$. The Poiseuille temperature carries a free additive constant, so $E_q$ is the gauge-invariant measure there.

\subsection{PGD and PGD--GSIS across transport regimes}\label{sec:pgd_cavity}
\begin{figure}[tbp]
\centering
\includegraphics[width=0.95\textwidth]{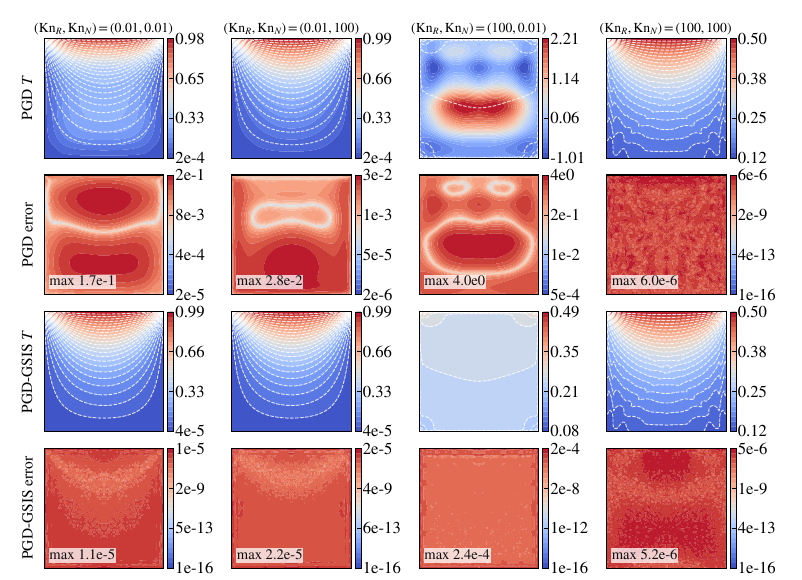}
\caption{Isothermal cavity at $(\Kn_R,\Kn_N)=(0.01,0.01)$, $(0.01,100)$, $(100,0.01)$, $(100,100)$ (columns). Rows 1--2: pure PGD, temperature and pointwise error $|T-T_{\FR}|/\max|T_{\FR}|$; rows 3--4: PGD--GSIS. White dashed lines are full-order contours. Each panel has its own color scale; the boxed number is the maximum error in the panel.}
\label{fig:comparison}
\end{figure}

Figure~\ref{fig:comparison} compares pure PGD and PGD--GSIS on the cavity. Pure PGD has maximum pointwise errors of $1.7\times10^{-1}$ at $(0.01,0.01)$, $2.8\times10^{-2}$ at $(0.01,100)$ and $4.0$ at $(100,0.01)$, where the temperature leaves the physical range $[0,1]$; of the four pairs shown, only the ballistic one is accurate ($6.0\times10^{-6}$). At the three failing pairs the spatial sweeps of pure PGD have margins $1-\rho_{\rm sw}\leq3\times10^{-4}$ (Table~\ref{tab:fourier}), so that even the full budget of $50$ sweeps per mode removes less than $2\%$ of the slow error components; at the ballistic pair the margin is close to one. The hydrodynamic failure is the most severe. There the slow component is the heat flux, which normal scattering nearly conserves, and enrichment increases the error from $0.63$ at rank one to $3.2$ at rank twenty (Fig.~\ref{fig:l2_modes}): new modes are fitted to an unconverged state instead of correcting it. PGD--GSIS reduces the maximum errors to $1.1\times10^{-5}$, $2.2\times10^{-5}$, $2.4\times10^{-4}$ and $5.2\times10^{-6}$. The recovered hydrodynamic field shows the expected physics: the bulk is nearly isothermal ($0.24\leq T\leq0.26$ away from the walls, against a global range $[0.08,0.49]$) and the temperature variation is confined to wall layers.

\begin{figure}[tbp]
\centering
\includegraphics[width=0.95\textwidth]{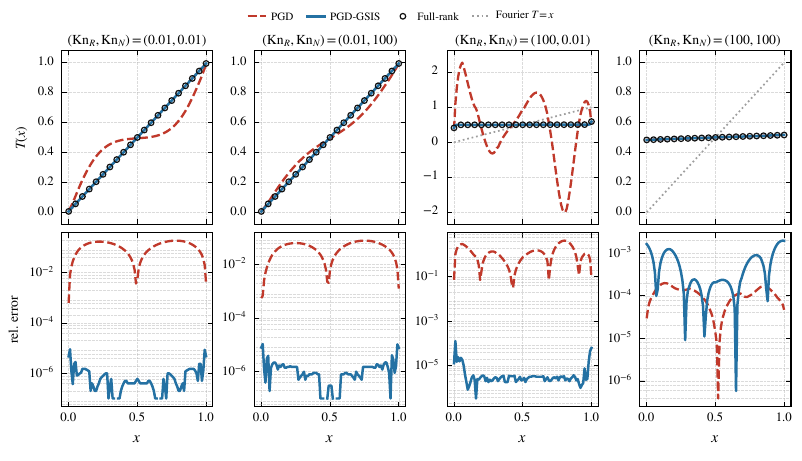}
\caption{Cross-plane conduction: $T$ on $y=H/2$ (top) and relative error against the full-order solution (bottom). Circles: full-order GSIS (``Full-rank'' in the legend); dashed: pure PGD; solid: PGD--GSIS; dotted: Fourier profile $T=x$.}
\label{fig:cross_comparison}
\end{figure}

\begin{figure}[tbp]
\centering
\includegraphics[width=0.95\textwidth]{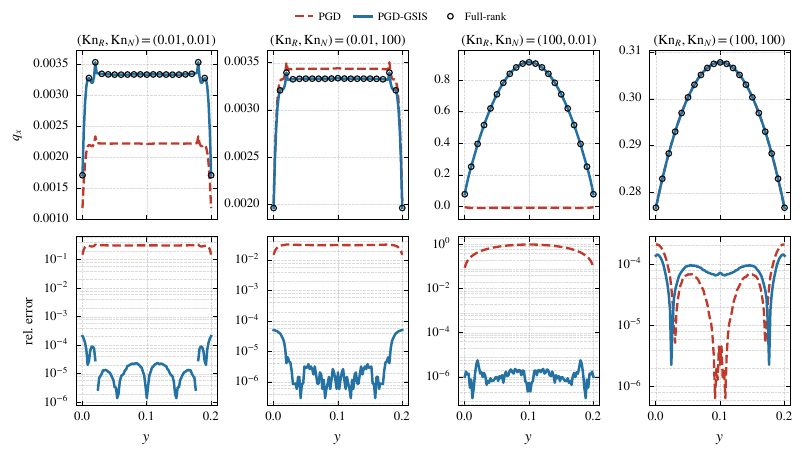}
\caption{Phonon Poiseuille flow: $q_x$ across the channel (top) and relative error against the full-order solution (bottom). Symbols as in Fig.~\ref{fig:cross_comparison}.}
\label{fig:poiseuille_comparison}
\end{figure}

The channel problems show the same dichotomy (Figs.~\ref{fig:cross_comparison} and~\ref{fig:poiseuille_comparison}), and PGD--GSIS lies on the full-order profiles in every regime. The profiles follow the mechanisms of Section~\ref{sec:tests}. In cross-plane conduction the diffusive profile is the Fourier line $T=x$; at $(100,0.01)$ and $(100,100)$ it is a plateau near $T=0.5$ with large wall jumps, since $\partial_xT=-3\delta_Rq_x$ is small whenever resistive scattering is rare, whether the flow is hydrodynamic or ballistic. In Poiseuille flow the diffusive profiles are flat at the Fourier value $\Kn_R/3=3.3\times10^{-3}$ with wall layers a few thousandths wide, of the order of $\ell_\GK$ and $\Kn_C$; at $(100,0.01)$, where $\ell_\GK=0.45>H$, the profile is parabolic with center value $0.90$, close to the no-slip Guyer--Krumhansl estimate $5H^2/(24\Kn_C)=0.83$; at $(100,100)$ it is a flattened arch between $0.277$ and $0.308$. With pure PGD, the flux of the diffusive Poiseuille pair $(0.01,0.01)$ is too small by a factor of about three and the hydrodynamic profiles are wrong at order one.

\subsection{Rank convergence and modal contributions}\label{sec:rank}
\begin{figure}[tbp]
\centering
\includegraphics[width=0.9\textwidth]{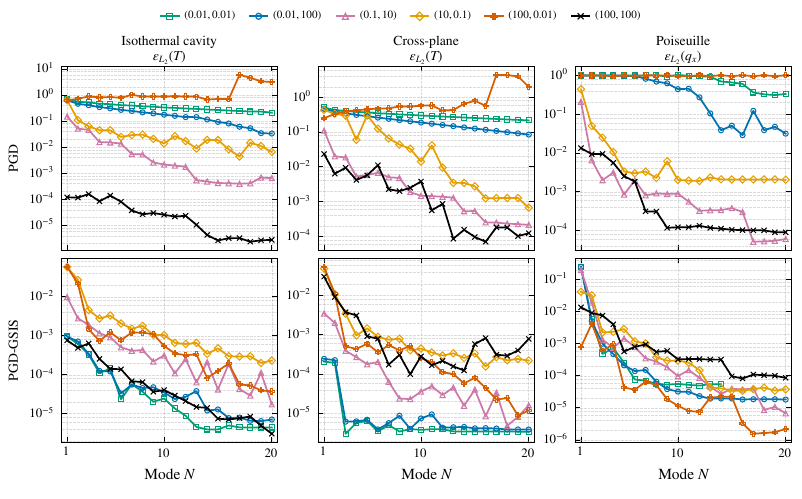}
\caption{Relative $L^2$ error versus retained rank for pure PGD (top) and PGD--GSIS (bottom), at the six Knudsen pairs. Columns: cavity $T$, cross-plane $T$, Poiseuille $q_x$ (profile at mid-length).}
\label{fig:l2_modes}
\end{figure}

Figure~\ref{fig:l2_modes} shows the error against the number of retained modes. With PGD--GSIS the diffusive pairs reach about $10^{-4}$ in the cavity and $7\times10^{-6}$ in the cross-plane problem by rank five; the hydrodynamic pairs end between $10^{-5}$ and $10^{-4}$ in all three problems; the ballistic pair ends at $3.1\times10^{-6}$ (cavity), $7.9\times10^{-4}$ (cross-plane) and $8.6\times10^{-5}$ (Poiseuille). With pure PGD the error stagnates or grows at the stalled pairs, as discussed above.

\begin{figure}[tbp]
\centering
\includegraphics[width=0.9\textwidth]{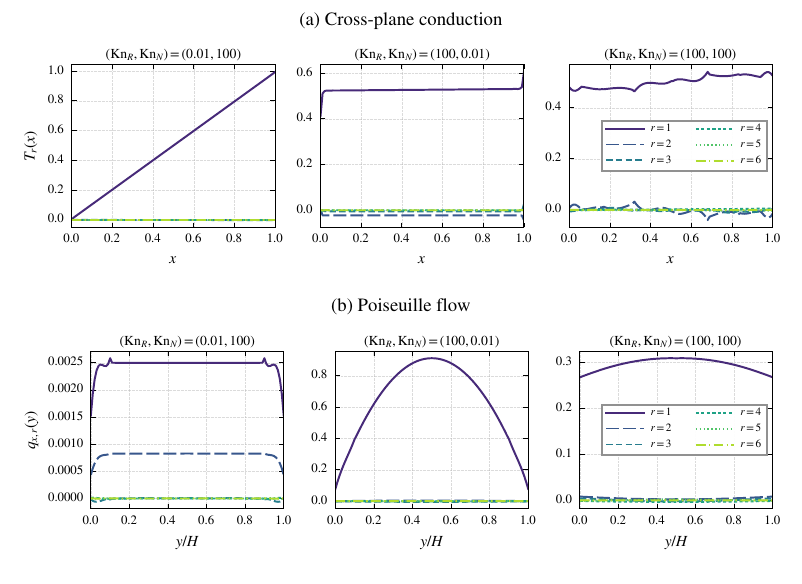}
\caption{First six PGD--GSIS macroscopic contributions in the channel problems. Top (a): cross-plane temperature contributions $T_r(x)=Y_r(x)\mathsf T_r$. Bottom (b): Poiseuille heat-flux contributions $q_{x,r}(y)$. Columns correspond to $(\Kn_R,\Kn_N)=(0.01,100)$, $(100,0.01)$ and $(100,100)$. Line styles identify $r=1,\ldots,6$; contributions are shown at their original magnitudes, with a separate vertical scale in each panel.}
\label{fig:modes_channels}
\end{figure}

Figure~\ref{fig:modes_channels} shows the first six PGD--GSIS macroscopic contributions for the channel problems. In cross-plane conduction the first temperature contribution carries the nearly linear diffusive profile and the plateaus of the hydrodynamic and ballistic pairs. The sum of modes two to six, relative to the rank-six temperature field in $L^2$ norm, is $2.5\times10^{-4}$ (diffusive), $5.4\times10^{-2}$ (hydrodynamic) and $3.0\times10^{-2}$ (ballistic). In Poiseuille flow the leading heat-flux contribution reproduces the hydrodynamic parabola, while the diffusive flux is shared by the first two modes. These plots illustrate how the constructed modes contribute to the macroscopic fields; the angular rank bound of Eq.~\eqref{eq:rank2} concerns the full distribution. The ranks at which enrichment stops (Table~\ref{tab:pgd_cost}) follow the ordering of Section~\ref{sec:regimes_rank}: three in the diffusive limit, four to seven in the ballistic limit, and most, up to fourteen, at $(10,0.1)$, where optical depths of order one to ten span the domain and wall layers are thick.

\begin{figure}[tbp]
\centering
\includegraphics[width=0.9\textwidth]{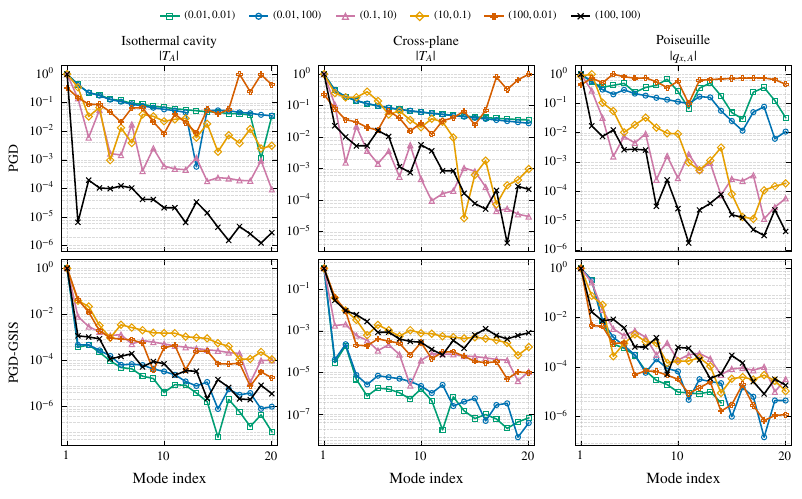}
\caption{Modal amplitudes during enrichment for pure PGD (top) and PGD--GSIS (bottom), with enrichment continued to twenty modes. Columns as in Fig.~\ref{fig:l2_modes}. The curves show the evolution of successive macroscopic contributions; only the decay along a curve is meaningful, and the amplitudes are not error bounds.}
\label{fig:amplitude_decay}
\end{figure}

The modal amplitudes in Fig.~\ref{fig:amplitude_decay} record the actual evolution of successive macroscopic contributions during enrichment and complement the reference-error curves of Fig.~\ref{fig:l2_modes}. They are the quantities monitored by the increment test~\eqref{eq:enrichment}; these diagnostic runs continue to twenty modes. For PGD--GSIS every curve falls by more than two decades over twenty modes; for pure PGD the hydrodynamic curve grows or stagnates. The amplitude is nevertheless not an error estimate: the ballistic Poiseuille pair decays more slowly than several others yet reaches a small error, because the separated terms are not orthogonal.

\subsection{Computational cost and memory scaling}\label{sec:cost}
\begin{figure}[tbp]
\centering
\includegraphics[width=\textwidth]{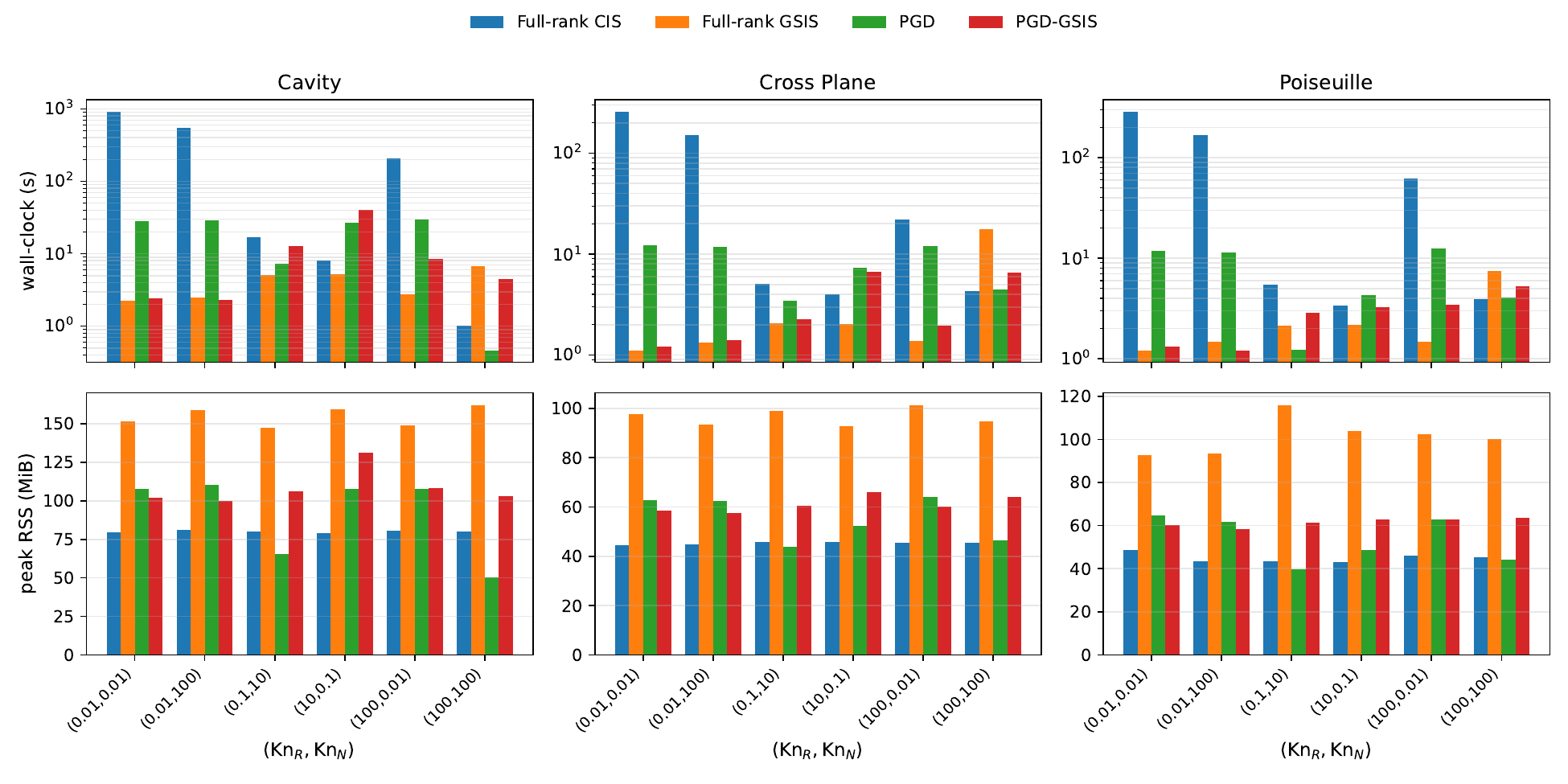}
\caption{Wall-clock time (top, logarithmic) and peak resident memory (bottom) of the four solvers at the six Knudsen pairs; $N_\theta\times N_\phi=80\times40$, $12$ threads. ``Full-rank'' in the legend denotes the full-order solvers.}
\label{fig:cost}
\end{figure}

Table~\ref{tab:pgd_cost} and Fig.~\ref{fig:cost} compare four solvers on all pairs with $12$ OpenMP threads: full-order source iteration (CIS), full-order GSIS, pure PGD, and PGD--GSIS with enrichment stopped at $\epsilon_{\rm en}=10^{-3}$.

\begin{table}[tbp]
\centering
\caption{Rank at which enrichment stops, with the average number of alternating pairs per mode in parentheses; $\dagger$: rank cap $M_{\max}=20$ reached without meeting $\epsilon_{\rm en}$. PGD denotes pure PGD (Section~\ref{sec:factor_eqs}).}
\label{tab:pgd_cost}
\small
\begin{tabular}{l l l l l l l}
\toprule
$(\Kn_R,\Kn_N)$ & \multicolumn{2}{c}{cavity} & \multicolumn{2}{c}{cross-plane} & \multicolumn{2}{c}{Poiseuille}\\
\cmidrule(lr){2-3}\cmidrule(lr){4-5}\cmidrule(lr){6-7}
 & PGD & PGD--GSIS & PGD & PGD--GSIS & PGD & PGD--GSIS\\
\midrule
  $(0.01,0.01)$ & $20^{\dagger}\,(9.8)$ & $3\,(5.0)$ & $20^{\dagger}\,(10.0)$ & $3\,(6.3)$ & $20^{\dagger}\,(9.8)$ & $3\,(6.3)$ \\
  $(0.01,100)$ & $20^{\dagger}\,(10.0)$ & $3\,(5.0)$ & $20^{\dagger}\,(10.0)$ & $3\,(7.0)$ & $20^{\dagger}\,(9.5)$ & $3\,(6.0)$ \\
  $(0.1,10)$ & $8\,(9.0)$ & $7\,(7.9)$ & $8\,(9.4)$ & $4\,(8.0)$ & $4\,(8.8)$ & $4\,(9.3)$ \\
  $(10,0.1)$ & $20^{\dagger}\,(9.5)$ & $14\,(8.1)$ & $14\,(10.0)$ & $8\,(8.1)$ & $11\,(9.0)$ & $5\,(7.8)$ \\
  $(100,0.01)$ & $20^{\dagger}\,(10.0)$ & $6\,(6.8)$ & $20^{\dagger}\,(10.0)$ & $4\,(7.0)$ & $20^{\dagger}\,(10.0)$ & $5\,(8.4)$ \\
  $(100,100)$ & $2\,(4.0)$ & $4\,(6.3)$ & $9\,(10.0)$ & $7\,(10.0)$ & $8\,(10.0)$ & $6\,(10.0)$ \\
\bottomrule
\end{tabular}
\end{table}

\begin{figure}[tbp]
\centering
\includegraphics[width=\textwidth]{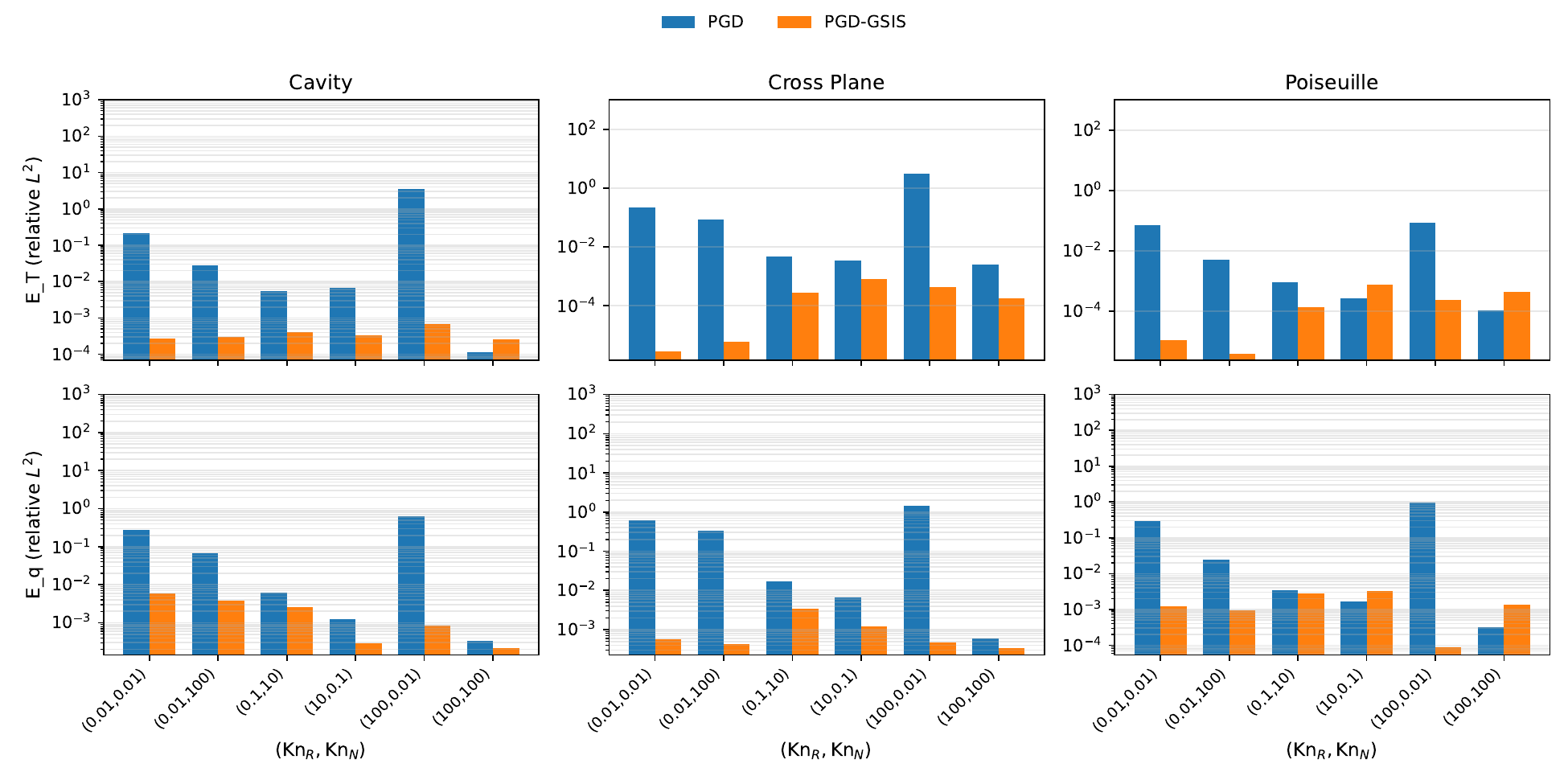}
\caption{Errors $E_T$ (top) and $E_q$ (bottom) of pure PGD and of PGD--GSIS against full-order GSIS. The Poiseuille $E_T$ is affected by the free temperature constant.}
\label{fig:accuracy}
\end{figure}

Full-order CIS needs $10^2$--$10^4$ sweeps and up to $911$\,s; its cost is largest where the margin $1-\rho(A_{\mathcal K})$ of Table~\ref{tab:fourier} is smallest. Full-order GSIS converges in $8$--$216$ iterations and $1.1$--$17.7$\,s. PGD--GSIS stops at ranks $3$--$14$ after $15$--$114$ synthetic updates in total---the same order as full-order GSIS, as the analysis of Section~\ref{sec:cis_fourier} predicts---and takes $1.2$--$40.4$\,s. Its heat-flux error stays below $6\times10^{-3}$ and its cavity temperature error below $7\times10^{-4}$ on every pair (Fig.~\ref{fig:accuracy}). At the diffusive pairs $E_q$ exceeds $E_T$ in all three problems, by more than an order of magnitude in the cavity, where these pairs also give the largest $E_q$; this is the $\|e\|/\Kn_R$ sensitivity of the kinetic flux discussed in Section~\ref{sec:param_error}. Pure PGD takes $0.5$--$29.8$\,s, but at the stalled pairs this is the cost of exhausting the budget of $200$ alternating pairs without converging ($E_q$ up to $1.5$), not a speedup. It matches PGD--GSIS only where the sweep margin $1-\rho_{\rm sw}$ is large.

\begin{figure}[tbp]
\centering
\includegraphics[width=0.8\textwidth]{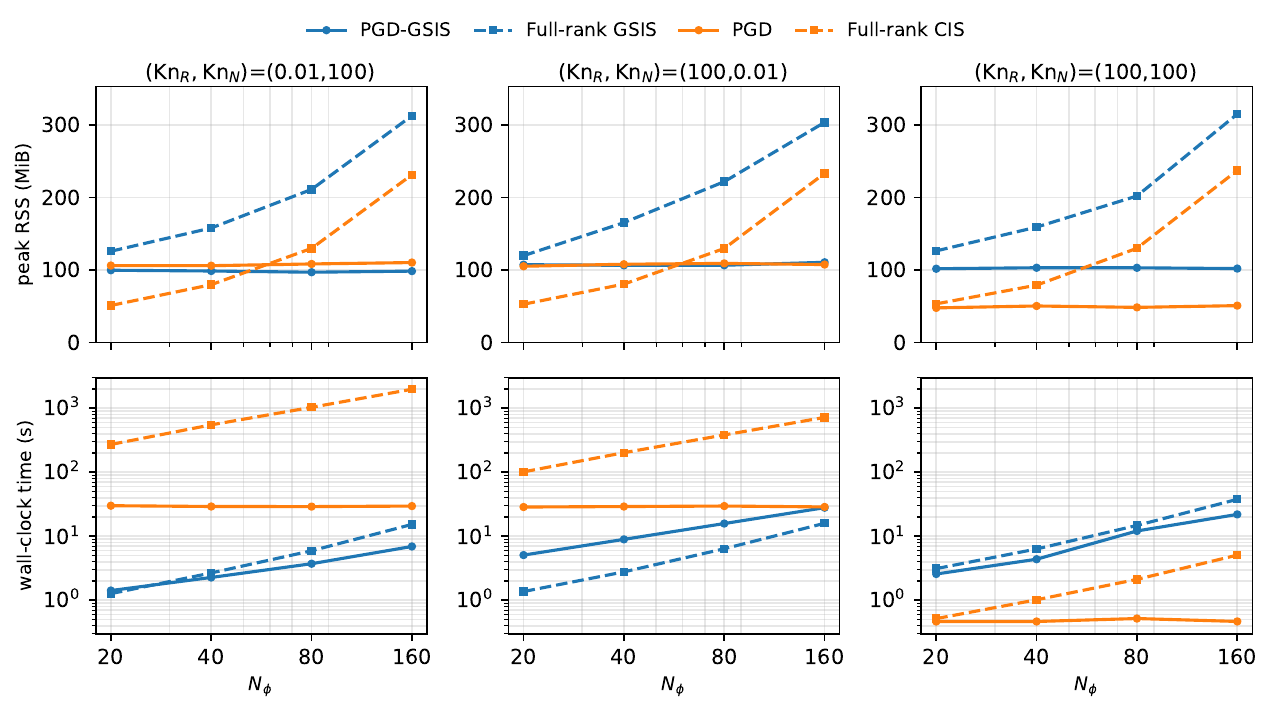}
\caption{Cavity: peak resident memory (top) and wall-clock time (bottom) versus the polar node count at $N_\theta=80$. Solid: separated representation (``PGD'' in the legend); dashed: full order (``Full-rank''); blue: with synthetic acceleration (GSIS, PGD--GSIS); orange: without (CIS, pure PGD).}
\label{fig:nazim}
\end{figure}

At $N_\phi=40$, PGD--GSIS is not faster than full-order GSIS: its per-update cost is comparable (Section~\ref{sec:repcost}), and neither solver is limited by the number of updates. Its peak memory ($58$--$131$\,MiB) is below that of full-order GSIS ($93$--$162$\,MiB) but above that of full-order CIS (roughly $45$--$80$\,MiB), which stores no synthetic operator: at this angular resolution the peak is dominated by arrays that do not depend on $N_\phi$. The difference appears when the polar resolution grows (Fig.~\ref{fig:nazim}). Full-order memory increases by about $1.2$\,MiB per polar node, the size of one polar slice of the distribution ($N_xN_\theta$ values), whereas the PGD memory is flat because refining $\phi$ only lengthens the polar factors. At $N_\phi=160$ PGD--GSIS uses $2.7$--$3.2$ times less memory than full-order GSIS, and pure PGD $2.1$--$4.7$ times less than full-order CIS. The full-order CIS time grows from $2.7\times10^2$\,s to $2.0\times10^3$\,s at $(0.01,100)$. The PGD--GSIS time still grows with $N_\phi$. In the implementation used here the high-order load and the wall fluxes are evaluated by full angular quadrature, which is the dominant $N_\phi$-dependent stage; the moment form of Section~\ref{sec:gsis} removes that dependence.

\subsection{Accuracy and cost of parametric PGD--GSIS}\label{sec:param_results}
\begin{figure}[tbp]
\centering
\includegraphics[width=0.80\textwidth]{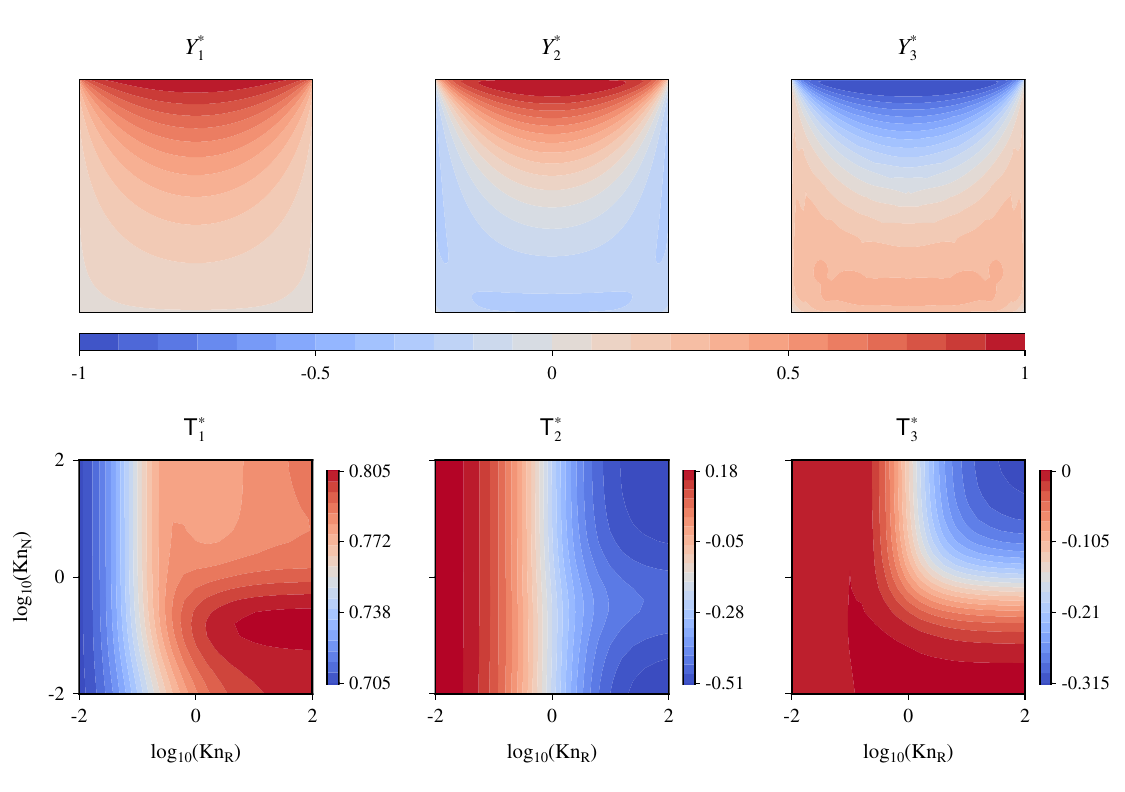}
\caption{First three temperature factor--coefficient pairs of the parametric cavity construction, with $r=1,2,3$ by column. Top: unit-peak spatial shapes $Y_r^*$ on the unit square, sharing the scale $[-1,1]$. Bottom: matching coefficients $\mathsf T_r^*$ over the Knudsen-number plane, each with its own linear color scale. Their products give the modal temperature contributions exactly; $|\mathsf T_r^*|$ measures their peak magnitude.}
\label{fig:parametric_temperature_modes}
\end{figure}

Each test problem carries one parametric construction over $\Kn_R,\Kn_N\in[10^{-2},10^2]$ ($21\times21$ nodes, rank cap $30$, tolerances $10^{-3}$, joint polar update). Figure~\ref{fig:parametric_temperature_modes} pairs the first three temperature spatial factors of the cavity construction with their parameter-dependent coefficients. To expose shape and magnitude separately, set $Y_r^*=Y_r/\max|Y_r|$ and $\mathsf T_r^*=\max|Y_r|\,\mathsf T_r$, so that $Y_r^*\mathsf T_r^*=Y_r\mathsf T_r$. The magnitude $|\mathsf T_r^*(\boldsymbol\delta)|$ is then the peak temperature contribution of mode $r$, in units of the imposed temperature difference. Separate unit-peak normalization of the shapes does not indicate their relative importance.

The first spatial factor is one-signed and peaks at the hot wall; its coefficient varies from about $0.70$ to $0.81$ across the parameter box. On the Fourier side the first two temperature contributions dominate, at about $0.70$ and $0.18$. The second coefficient changes sign toward the kinetic side and reaches about $-0.5$ at $\Kn_R=10^2$, while the third grows to about $-0.3$ only where both Knudsen numbers are large. These pairs illustrate how one shared spatial basis adjusts its contributions across regimes. Individual PGD modes depend on the construction and do not have a unique physical interpretation.

\begin{figure}[tbp]
\centering
\includegraphics[width=0.70\textwidth]{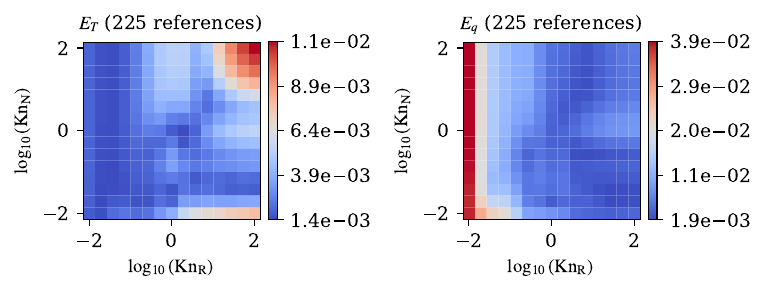}
\caption{Parametric cavity reconstruction at rank $30$: relative error of $T$ (left) and of $|\bm q|$ (right) against $225$ full-order GSIS solutions, as discrete $\ell^2$ ratios on a $101\times101$ sampling of the field. The color scales differ: $1.4\times10^{-3}$--$1.1\times10^{-2}$ (left) and $1.9\times10^{-3}$--$3.9\times10^{-2}$ (right).}
\label{fig:error_map}
\end{figure}

The reconstruction is scored against $225$ independent full-order GSIS solutions on a $15\times15$ lattice spanning the same box. Most lattice points lie between stored nodes, so the scores combine representation and interpolation errors. At rank $30$ the cavity temperature error lies between $0.14\%$ and $1.15\%$ and the heat-flux error between $0.19\%$ and $3.87\%$ (Fig.~\ref{fig:error_map}, Table~\ref{tab:parametric}); both channel constructions are more accurate. The error distribution follows Section~\ref{sec:param_error}. The largest $E_q$ lie along the edge $\Kn_R=10^{-2}$, where the flux is $\mathcal O(\Kn_R)$ and a uniform distribution error becomes a large relative flux error; the largest $E_T$ ($1.1\times10^{-2}$) lies at the ballistic corner, where the synthetic temperature is most sensitive to the representation. At the four corner pairs, scoring against the fixed-parameter PGD--GSIS solutions instead of full-order ones changes the result only in the fourth significant figure, since those solutions are far more accurate than the parametric reconstruction.

\begin{table}[tbp]
\centering
\caption{Parametric constructions ($21\times21$ nodes): smallest/largest relative error (\%) over the $225$ full-order references at three truncation ranks of the same stored factors. $E_q$ is the error of $|\bm q|$ for the cavity and of $q_x$ for the channels; the Poiseuille $E_T$ is computed after removing the mean.}
\label{tab:parametric}
\small
\begin{tabular}{l rrr rrr}
\toprule
Case & \multicolumn{3}{c}{$E_T$ (\%) at rank} & \multicolumn{3}{c}{$E_q$ (\%) at rank}\\
\cmidrule(lr){2-4}\cmidrule(lr){5-7}
 & $10$ & $20$ & $30$ & $10$ & $20$ & $30$\\
\midrule
cavity & $1.17/4.63$ & $0.22/2.10$ & $0.14/1.15$ & $0.94/35.1$ & $0.35/9.09$ & $0.19/3.87$\\
cross-plane & $0.23/4.60$ & $0.029/0.71$ & $0.013/0.27$ & $0.32/21.9$ & $0.063/4.44$ & $0.016/2.60$\\
Poiseuille & $0.029/1.89$ & $0.014/0.25$ & $0.0030/0.056$ & $0.15/33.2$ & $0.062/4.15$ & $0.015/2.62$\\
\bottomrule
\end{tabular}
\end{table}

These errors are one to two orders of magnitude above the fixed-parameter errors at the same pairs, and Table~\ref{tab:parametric} shows why they should not be read as the limit of the method: from rank $10$ to $30$ the largest cavity errors fall by factors of $4.0$ ($E_T$) and $9.1$ ($E_q$), the largest Poiseuille $E_q$ by $12.7$, and the contribution of the thirtieth mode is still $0.68\%$ (cavity), $0.15\%$ (cross-plane) and $0.04\%$ (Poiseuille) of the first. Every construction stopped at the rank cap. The larger ranks are consistent with the shared-basis requirement of Section~\ref{sec:param_error}: one space must approximate the temperature and flux fields across the whole parameter box.

Interpolation between nodes was assessed by withholding every other node in each direction: the coefficients $\mathsf T_r$ and $\mathsf Q_r$ on the remaining $11\times11$ nodes were interpolated linearly in $(\log_{10}\Kn_R,\log_{10}\Kn_N)$ to the $320$ withheld nodes and compared with the values the construction stores there. The median errors are $2.7\times10^{-3}$ (temperature) and $4.9\times10^{-3}$ (heat flux), the maxima $1.0\times10^{-2}$ and $6.9\times10^{-2}$. Since the spacing is doubled and linear interpolation errs quadratically in the spacing, the off-grid error of the full $21\times21$ table is expected to be about four times smaller wherever the coefficients are smooth.

The three constructions take $1.26$\,min (cavity), $0.52$\,min (cross-plane) and $0.60$\,min (Poiseuille) on $32$ threads, with $9$--$10$ alternating pairs per mode. Peak memory is $12.0$\,GiB for the cavity and $4.9$\,GiB for the channels, dominated by the per-node HDG operators of Section~\ref{sec:param_cost}.

\paragraph{Role of the joint polar update} Without the joint update, the synthetic refresh is unstable at the rarefied corner. The $21\times21$ construction then diverged at $(\Kn_R,\Kn_N)=(63,100)$, $(100,63)$ and $(100,100)$, where the temperature left $[0,1]$ by up to $4.2$, and on a $5\times5$ grid additional modes did not cure the instability ($E_T=0.93$ at $(100,100)$ with $60$ modes). With the joint update the iteration remained stable over the whole box. Besides the $21\times21$ constructions above, $5\times5$ constructions with $30$ modes were scored at the six fixed-parameter pairs, against full-order solutions, and at eight transition-regime nodes, against converged single-node PGD--GSIS solutions. Their largest errors are $E_T=4.0\times10^{-3}$ and $E_q=5.3\times10^{-3}$ in cross-plane conduction and $E_T=9.0\times10^{-3}$ and $E_q=2.4\times10^{-2}$ in the cavity, with the largest $E_T$ again at $(100,100)$. The update adds only $m\times m$ solves per polar and parameter node: on a $21$-node line at $\Kn_R=10$ with $\Kn_N\in[10^{-2},10^2]$ and rank cap $25$, it reduced the largest residual of the frozen-source transport problem from $0.46$ to $0.012$ at essentially unchanged time ($4.2$ versus $4.1$\,min).

\section{Conclusions}\label{sec:conclusions}
We have combined proper generalized decomposition with the general synthetic iterative scheme for the steady linearized gray Callaway model on two-dimensional domains. Three findings organize the results.

First, the splitting of the distribution into a space--azimuth factor and a polar factor is intrinsically low-rank. Upwinding depends on the azimuth alone, and along each characteristic the polar angle enters through the stretching of the in-plane optical depth. The kernel $\mathrm{e}^{-\tau/\sin\phi}$ explains the small polar ranks, and polynomial angular states of degree $N$ have rank at most $N+1$. Leading equilibria have rank at most two, while their first streaming correction can require rank three; wall layers add further angular structure.

Second, the treatment of the macroscopic coupling decides convergence. Pure PGD keeps the scattering source implicit in the polar update, but its spatial update moves macroscopic information only through sweeps that lag the moments of the current mode. A Fourier analysis shows that these sweeps have the vanishing contraction margin of source iteration, and the reported pure-PGD experiments stall or grow at the collision-dominated pairs. PGD--GSIS freezes the source within each alternating pair and refreshes it by the synthetic equations; its error propagation splits into the full-order GSIS operator and the image of the representation error, which in the periodic bulk is read only through degree-three angular moments, with a gain that is small in the Fourier regime. The conditional AP analysis shows that smooth states with controlled projected residuals, polar approximation errors and source increments reconstruct the first streaming correction, suppress those moments and match the synthetic fields to $o(\Kn_C)$. This preserves the leading Fourier flux and the viscous GK balance without identifying low-rank representability with an AP certificate. PGD--GSIS converged in a number of synthetic updates comparable to full-order GSIS and reproduced full-order solutions at ranks $3$--$14$, with heat-flux errors below $6\times10^{-3}$, on all three benchmarks from the diffusive to the ballistic regime. In the ballistic limit the kinetic refresh is faster. At the angular resolution of the benchmarks the wall time is comparable to full-order GSIS; the benefit is the weak dependence of separated storage on the polar resolution.

Third, a single spatial basis can represent four decades of both Knudsen numbers, provided all polar factors are updated jointly. Without this update, polar factors fitted to an earlier source leave representation errors that the synthetic map amplifies at large $\Kn_C$, and the construction diverged at the rarefied corner. The joint update costs one $m\times m$ system per polar and parameter node and stabilized the iteration over the whole box at essentially unchanged cost. The rank must be large enough to approximate the entire macroscopic solution family at the required accuracy. At rank $30$ the parametric cavity reconstruction was accurate to about $1\%$ in temperature and $4\%$ in heat flux and was still improving. Its error concentrates where the analysis predicts: in the relative heat flux at small $\Kn_R$, where the flux is $\mathcal O(\Kn_R)$, and in the temperature at the ballistic corner, where the synthetic temperature responds to representation errors with gain $\mathcal O(\Kn_C)$.

These observations suggest the next steps. A representation of the deviation from the Callaway equilibrium, as in micro--macro low-rank schemes~\cite{einkemmer2021collision}, could reduce relative-flux sensitivity. Reporting the synthetic flux together with its matching defect is another option in the Fourier regime. The per-node synthetic operators, which dominate the parametric memory, could be assembled on the fly. A further direction is to introduce separate PGD factors for the two Knudsen numbers and extend this separation to the macroscopic HDG equations. Shared spatial systems could then replace per-node factorizations and solves, potentially yielding substantial reductions in memory and computational cost. Uniform attainment of the AP residual conditions with fixed rank and iteration budgets, and a fully discrete AP analysis including kinetic layers, remain open. Nongray scattering, where frequency is a natural additional coordinate, and three-dimensional geometries, where the upwind direction depends on the polar angle, are the main extensions.

\section*{CRediT authorship contribution statement}
\textbf{Jia Liu:} Conceptualization, Methodology, Software, Validation, Formal analysis, Investigation, Visualization, Writing -- original draft.\\
\textbf{Wei Su:} Conceptualization, Methodology, Supervision, Funding acquisition, Writing -- review \& editing.

\section*{Declaration of competing interest}
The authors declare that they have no known competing financial interests or personal relationships that could have appeared to influence the work reported in this paper.

\section*{Data availability}
The code and data that support the findings of this study are available from the corresponding author upon reasonable request.

\section*{Declaration of generative AI and AI-assisted technologies in the manuscript preparation process}
During the preparation of this work the authors used Claude (Anthropic) to edit the language and structure of the manuscript and to check derivations. After using this tool, the authors reviewed and edited the content as needed and take full responsibility for the content of the published article.


\appendix
\section{Asymptotic limits of PGD--GSIS}\label{app:limits}
We analyze the frozen-source construction of Algorithm~\ref{alg:pgd} in three limits. In the diffusive and hydrodynamic limits the collision term is stiff, and the question is asymptotic preservation. In the ballistic limit the question is whether the synthetic coupling degrades free streaming. As in the continuous-coordinate PGD analysis of~\cite{yin2026asymptotic}, spatial and angular discretization errors are excluded. Sections~\ref{app:ap_formal}--\ref{app:ap_matching} develop the estimates common to the two collision-dominated limits, Sections~\ref{app:diffusive}--\ref{app:ballistic} treat the three limits, and Section~\ref{app:ap_scope} states the scope. The collision-dominated results concern smooth bulk fields on a fixed subdomain $\mathcal D_0\Subset\mathcal D$, or smooth periodic fields. They give sufficient, measurable conditions on the retained polar space, the projected transport residual and the source increment; they do not assume that an increment test or a fixed iteration cap satisfies these conditions, and do not assert uniform resolution of kinetic layers.

\subsection{Formal expansion of the frozen-source pair}\label{app:ap_formal}
As in~\cite{yin2026asymptotic}, the mechanism can first be seen from a formal expansion of the iteration. Set $\lambda=\Kn_C$ and define the streaming operator and frozen equilibrium by
\begin{equation}
 Df=\bm\Omega\cdot\nabla f,\qquad
 E_n=\frac{T_s^n+3\chi\bm\Omega\cdot\bm q_s^n}{4\pi}.
 \label{eq:ap_equilibrium}
\end{equation}
Away from walls, Eqs.~\eqref{eq:polar} and~\eqref{eq:spatial} of PGD--GSIS, divided by $\delta_C$, are the two partial Galerkin conditions of the frozen-source problem $I+\lambda DI=E_n$. With $R=I_m+\lambda DI_m-E_n$, they read
\begin{equation}
 (X_m,R)_{x\theta}=0\quad\text{for every }\phi,\qquad
 \int_0^\pi\Phi_mR\dd\phi=0\quad\text{for every }(\bm x,\theta).
 \label{eq:ap_pair_rows}
\end{equation}
At $\lambda=0$ these are the normal equations of $\min\|E_n-I_{m-1}-X_m\Phi_m\|$ with respect to $\Phi_m$ and to $X_m$ separately. To leading order, the pair is therefore an alternating least-squares iteration for the best rank-one approximation of $E_n-I_{m-1}$, that is, a power iteration for its leading singular pair, and greedy enrichment at a frozen source is singular-value deflation. Its contraction factor per pair is the squared ratio of the two largest singular values of $E_n-I_{m-1}$ and does not depend on $\lambda$, whereas the spatial sweeps of pure PGD have margin $\mathcal O(\lambda^2)$ (Section~\ref{sec:cis_fourier}). By Proposition~\ref{prop:pn}, $E_n$ has polar factors in $G_1=\operatorname{span}\{1,\sin\phi\}$, so two converged modes reproduce it.

Next expand $I_m=I^{(0)}+\lambda I^{(1)}+\mathcal O(\lambda^2)$ and suppose that $R$ vanishes order by order. Then $I^{(0)}=E_n$ and $I^{(1)}=-DE_n$, with
\begin{equation}
 DE_n=\frac{\sin\phi\,\bm s\cdot\nabla T_s^n+3\chi\sin^2\phi\,\bm s\cdot\nabla(\bm s\cdot\bm q_s^n)}{4\pi}.
 \label{eq:ap_first_order}
\end{equation}
The state $E_n-\lambda DE_n$ is a polynomial of degree two in $\bm\Omega$. Its polar factors lie in $G_2=\operatorname{span}\{1,\sin\phi,\sin^2\phi\}$, so it has rank three, and the degree-three kernel $K_{ijk}$ is orthogonal to it. The high-order term of the synthetic system therefore enters only at $\mathcal O(\lambda^2)$, the kinetic moments match the synthetic fields as in Eq.~\eqref{eq:ap_moment_expansion} below, and the synthetic system reduces to the Fourier or GK balance of Sections~\ref{app:diffusive} and~\ref{app:hydro}.

This argument assumes three things that the algorithm does not guarantee. First, Eq.~\eqref{eq:ap_pair_rows} enforces only the rows of mode $m$; for $R$ to vanish, the rows of the frozen earlier modes must hold as well. Second, the source is refreshed after every pair, so the earlier modes were computed with earlier sources. Third, an expansion of the factors, rather than of the state, requires simple leading singular values; otherwise the rank-one factors are not unique, as can happen where singular values cross in the parameter box.

The first assumption matters even at a frozen source. We ran the iteration of \ref{app:fourier} with $\bm U_s$ frozen, $k=\pi$ and $\Kn_R=\Kn_N=\varepsilon$, so that $\lambda=\varepsilon/2$, and alternated every mode to an increment below $10^{-11}$. At $\varepsilon=10^{-2}$ and $10^{-3}$ the first mode contracted by $3.4\times10^{-2}$ and $3.5\times10^{-2}$ per pair, against $(\sigma_2/\sigma_1)^2=3.5\times10^{-2}$, and no mode needed more than nine pairs. Two modes left a relative $L^2$ error of $7.9\times10^{-4}$ and $8.0\times10^{-5}$, of order $\lambda$. Three modes already spanned $G_2$ to within $\zeta_n\leq7\times10^{-7}$ (relative, in $L^2$), yet the error remained of order $\lambda$, at $1.4\times10^{-4}$ and $1.4\times10^{-5}$. To three digits it equaled the residual indicator $\eta_n$ of Eq.~\eqref{eq:ap_indicators}: the rows of the frozen modes held only to $\mathcal O(\lambda)$. The fourth mode reduced the error to $4.0\times10^{-6}$ and $4.1\times10^{-8}$, of order $\lambda^2$, and the degree-three content followed the same orders. The normal-dominated scaling ($\Kn_R=1$, $\Kn_N=\varepsilon$) gave the same orders. The greedy construction thus needs more modes than the rank of the Chapman--Enskog state, and of the quantities above only $\eta_n$ detects this; neither $\zeta_n$ nor the inner increments do. Sections~\ref{app:ap_conditions}--\ref{app:hydro} replace the three assumptions by measurable indicators and prove the corresponding estimates.

\subsection{Retained-mode indicators}\label{app:ap_conditions}
With $\lambda$, $D$ and $E_n$ as in Eq.~\eqref{eq:ap_equilibrium}, let $I_m^n$ be the state after the polar/spatial pair with source $\bm U_s^n$, and let $I_-^n$ be the state used for the preceding synthetic refresh. Thus
\begin{equation}
 \bm U_s^n=\mathcal G[I_-^n],\qquad \bm U_s^{n+1}=\mathcal G[I_m^n].
 \label{eq:ap_source_indices}
\end{equation}
The analysis applies after a synthetic refresh, with the same temperature gauge. Let $V_n=\operatorname{span}\{\Phi_1,\ldots,\Phi_m\}\subset L^2(0,\pi;\dd\phi)$ and let $P_n$ be its orthogonal projector, acting at each $(\bm x,\theta)$. The norm $\|\cdot\|_s$ is $H^s(\mathcal D_0;L^2(\dd\theta\dd\phi))$ for angular states and $H^s(\mathcal D_0)$ for macroscopic fields. Since the polar factors are independent of $\bm x$, $P_n$ commutes with spatial derivatives and has norm at most one in these spaces; it need not commute with $D$.

The collision-scaled residual and its projection are
\begin{equation}
 R_n=I_m^n+\lambda DI_m^n-E_n,\qquad C_n=P_nR_n.
 \label{eq:ap_projected_residual}
\end{equation}
These are diagnostics of the existing iteration. The continuous-coordinate version of Eq.~\eqref{eq:spatial}, multiplied by $\lambda$, enforces the row $\int\Phi_mR_n\dd\phi=0$ in the bulk when the spatial solve is exact. The frozen earlier modes do not enforce the other rows. If $\{\psi_j\}$ is an orthonormal basis of $V_n$, then
\begin{equation}
 C_n=\sum_{j=1}^{\dim V_n}\psi_j\int_0^\pi\psi_jR_n\dd\phi.
 \label{eq:ap_residual_rows}
\end{equation}
Every row uses polar contractions of the retained modes and their streaming derivatives. This avoids inferring the complete residual from a single Galerkin condition. Finite spatial-sweep errors are included in $C_n$.

Define the residual, approximation and source-increment indicators
\begin{equation}
 \begin{aligned}
 \eta_n&=\|C_n\|_3,\\
 \zeta_n&=\|(1-P_n)E_n\|_3+\lambda\|(1-P_n)DE_n\|_2,\\
 \omega_n&=\|\bm U_s^{n+1}-\bm U_s^n\|_2,
 \qquad \mathfrak a_n=\eta_n+\zeta_n+\omega_n.
 \end{aligned}
 \label{eq:ap_indicators}
\end{equation}
The quantity $\zeta_n$ checks whether the current polar space represents the equilibrium and its first streaming correction. By Proposition~\ref{prop:pn}, both lie in $G_2=\operatorname{span}\{1,\sin\phi,\sin^2\phi\}$, so $G_2\subset V_n$ is sufficient for $\zeta_n=0$. This is a sufficient condition on the space, not a claim that three greedy modes automatically span $G_2$. In the diffusive CE scaling, if the spatial derivatives of $\bm q_s$ are also $\mathcal O(\varepsilon)$, $G_1=\operatorname{span}\{1,\sin\phi\}\subset V_n$ already gives $\zeta_n=\mathcal O(\varepsilon^2)$.

\subsection{Bulk reconstruction and matching}\label{app:ap_matching}
\begin{proposition}[Bulk reconstruction and matching]\label{prop:ap}
Let $\lambda=\Kn_C\to0$, with $\lambda\asymp\varepsilon$ in the scalings of Sections~\ref{app:diffusive} and~\ref{app:hydro}. Assume exact angular integrals and synthetic solves, and uniform bulk bounds
\begin{equation}
 \|I_m^n\|_4+\|I_-^n\|_4+\|\bm U_s^n\|_4+\|\bm U_s^{n+1}\|_4\leq C,
 \label{eq:ap_regularity}
\end{equation}
with $C$ independent of $\varepsilon$. Then the actual retained state satisfies
\begin{equation}
 \|I_m^n-E_n+\lambda DE_n\|_2\leq C(\lambda^2+\eta_n+\zeta_n),
 \label{eq:ap_reconstruction}
\end{equation}
and its kinetic moments match the refreshed synthetic fields with
\begin{equation}
 \|\mathcal K[I_m^n]-\bm U_s^{n+1}\|_2\leq C(\lambda^2+\mathfrak a_n).
 \label{eq:ap_matching}
\end{equation}
Moreover,
\begin{equation}
 \|\bm M[I_m^n]\|_2\leq C(\lambda^2+\eta_n+\zeta_n),\qquad
 \|\lambda\nabla\cdot\bm H[I_m^n]\|_0\leq C\lambda(\lambda^2+\eta_n+\zeta_n).
 \label{eq:ap_hot_bound}
\end{equation}
No bound on the number of pairs needed to make the indicators small is asserted.
\end{proposition}
\begin{proofof}
Since $P_nI_m^n=I_m^n$, Eq.~\eqref{eq:ap_projected_residual} gives the exact identity
\begin{equation}
 I_m^n=P_nE_n-\lambda P_nDI_m^n+C_n.
 \label{eq:ap_projected_identity}
\end{equation}
Write $d_n=(1-P_n)E_n$. Applying $D$ to this identity and substituting once more yields
\begin{equation}
 \begin{split}
 I_m^n-E_n+\lambda DE_n={}&-d_n+\lambda(1-P_n)DE_n+\lambda P_nDd_n\\
 &+\lambda^2P_nDP_nDI_m^n-\lambda P_nDC_n+C_n.
 \end{split}
 \label{eq:ap_expansion_identity}
\end{equation}
Boundedness of $D:H^{s+1}\to H^s$, $\|P_n\|\leq1$, and Eq.~\eqref{eq:ap_regularity} prove~\eqref{eq:ap_reconstruction}. This step uses the residual rows of the spatial update, without deducing a vector identity from one scalar projection.

The equilibrium is a $P_1$ state and $DE_n$ a polynomial angular state of degree at most two. Orthogonality of the degree-three kernel $K_{ijk}$ to both therefore gives~\eqref{eq:ap_hot_bound}, including the two spatial derivatives in $\nabla\cdot\bm H$. Taking energy and flux moments of~\eqref{eq:ap_reconstruction} gives
\begin{equation}
 \begin{aligned}
 T_m^{\kin,n}&=T_s^n-\lambda\chi\nabla\cdot\bm q_s^n+\mathcal O(\lambda^2+\eta_n+\zeta_n),\\
 \bm q_m^{\kin,n}&=\chi\bm q_s^n-\frac{\lambda}{3}\nabla T_s^n+\mathcal O(\lambda^2+\eta_n+\zeta_n),
 \end{aligned}
 \label{eq:ap_moment_expansion}
\end{equation}
where the remainders are in $H^2(\mathcal D_0)$. By~\eqref{eq:ap_source_indices}, $\nabla\cdot\bm q_s^n=0$ and
\begin{equation}
 \frac13\nabla T_s^n+\delta_R\bm q_s^n
 =\lambda\left[\frac15\nabla^2\bm q_s^n+\nabla\cdot\bm H[I_-^n]\right].
 \label{eq:ap_previous_balance}
\end{equation}
Using $1-\chi=\lambda\delta_R$, the flux difference before the refresh is
\begin{equation}
 \bm q_m^{\kin,n}-\bm q_s^n
 =-\lambda^2\left[\frac15\nabla^2\bm q_s^n+\nabla\cdot\bm H[I_-^n]\right]
 +\mathcal O(\lambda^2+\eta_n+\zeta_n).
 \label{eq:ap_flux_matching_identity}
\end{equation}
The bracket is uniformly bounded in $H^2$ by~\eqref{eq:ap_regularity}. Adding the source increment proves~\eqref{eq:ap_matching}. In particular, matching is derived from the frozen-source reconstruction and the preceding synthetic balance; it is not assumed as a separate vanishing defect.
\end{proofof}

\subsection{Diffusive limit}\label{app:diffusive}
The diffusive scaling is $\Kn_R=\varepsilon\bar r$ and $\Kn_N=\varepsilon\bar n$, with fixed positive $\bar r,\bar n$. Then $\Kn_C=\varepsilon\bar c$, $\bar c=\bar r\bar n/(\bar r+\bar n)$, and $\chi=\bar r/(\bar r+\bar n)<1$. The leading collision balance gives $I^{(0)}=T^{(0)}/(4\pi)+3\chi\bm\Omega\cdot\bm q^{(0)}/(4\pi)$. Its flux moment is $\bm q^{(0)}=\chi\bm q^{(0)}$, hence $\bm q^{(0)}=0$. Taking the first moment at the next order gives
\begin{equation}
 \bm q=-\frac{\Kn_R}{3}\nabla T+\mathcal O(\varepsilon^3),\qquad \nabla\cdot\bm q=0.
 \label{eq:fourier}
\end{equation}
The limiting bulk temperature is therefore harmonic. The coefficient is $\Kn_R/3$, rather than $\Kn_C/3$, because the flux is relaxed only by resistive scattering.

\begin{corollary}[Diffusive limit]\label{cor:ap_diffusive}
In the diffusive scaling, let the assumptions of Proposition~\ref{prop:ap} hold and let $\mathfrak a_n=\mathcal O(\varepsilon^{1+\sigma_{\rm AP}})$ for some $0<\sigma_{\rm AP}\leq1$. Then, in $L^2(\mathcal D_0)$,
\begin{equation}
 \bm q_s^{n+1}=-\frac{\Kn_R}{3}\nabla T_s^{n+1}+\mathcal O(\varepsilon^3),\qquad
 \bm q_m^{\kin,n}=-\frac{\Kn_R}{3}\nabla T_s^{n+1}+\mathcal O(\varepsilon^{1+\sigma_{\rm AP}}).
 \label{eq:ap_fourier_matching}
\end{equation}
The error in the kinetic heat flux is therefore $o(\varepsilon)$, the limit of $\bm q_m^{\kin,n}/\varepsilon$ has the correct coefficient $-\bar r/3$, and every regular limiting temperature is harmonic in the bulk.
\end{corollary}
\begin{proofof}
The refreshed synthetic momentum equation, in $H^2$, implies $\bm q_s^{n+1}=\mathcal O(\varepsilon)$. Multiplying it by $\Kn_R$ gives the first relation in~\eqref{eq:ap_fourier_matching}, and Eq.~\eqref{eq:ap_matching} with $\lambda^2=\mathcal O(\varepsilon^{1+\sigma_{\rm AP}})$ gives the second. The exact synthetic energy balance makes every regular limiting temperature harmonic in the bulk.
\end{proofof}

\subsection{Hydrodynamic limit}\label{app:hydro}
The normal-dominated scaling is $\Kn_N=\varepsilon\bar n$ at fixed $\Kn_R$. Then $\Kn_C=\mathcal O(\varepsilon)$ and $\chi\to1$. The leading collision kernel consists of displaced equilibria $[T+3\bm\Omega\cdot\bm q]/(4\pi)$; the flux is undetermined at this order. With smooth fields, the next angular correction produces $N_{ij}=-(2/5)\Kn_C\partial_{\langle j}q_{i\rangle}+\mathcal O(\varepsilon^2)$ and the GK balance~\eqref{eq:gk}. At fixed $\Kn_R$, $\ell_\GK\to0$, so its strict bulk limit is Fourier conduction. To retain viscous transport at leading order, use the distinguished scaling
\begin{equation}
 \Kn_N=\varepsilon\bar n,\qquad \delta_R=\varepsilon\bar d_R,\qquad
 T=T_\ast+\varepsilon p_\GK,\qquad \bm q=\mathcal O(1),
 \label{eq:ap_gk_scaling}
\end{equation}
where $T_\ast$ is spatially constant and $\bar d_R\geq0$. In this scaling $\Kn_C/\varepsilon\to\bar n$ and the nontrivial bulk limit is
\begin{equation}
 \frac13\nabla p_\GK+\bar d_R\bm q-\frac{\bar n}{5}\nabla^2\bm q=0,
 \qquad \nabla\cdot\bm q=0.
 \label{eq:ap_gk_limit}
\end{equation}
This distinction separates a first-order GK correction from a limiting Brinkman/Stokes problem. Corollary~\ref{cor:ap_hydro} covers both.

\begin{corollary}[Hydrodynamic limit]\label{cor:ap_hydro}
Let the assumptions of Proposition~\ref{prop:ap} hold and let $\mathfrak a_n=\mathcal O(\varepsilon^{1+\sigma_{\rm AP}})$ for some $0<\sigma_{\rm AP}\leq1$. In the normal-dominated scaling, the synthetic fields and the kinetic moments satisfy the GK balance~\eqref{eq:gk} through its viscous order. In the distinguished scaling~\eqref{eq:ap_gk_scaling}, if the scaled temperature $p_\GK$ is also uniformly bounded in the norms of~\eqref{eq:ap_regularity}, regular limits of both sets of fields satisfy Eq.~\eqref{eq:ap_gk_limit}.
\end{corollary}
\begin{proofof}
In the fixed-$\Kn_R$ normal-dominated scaling, the refreshed momentum equation has high-order forcing $\mathcal O(\varepsilon^3+\varepsilon^{2+\sigma_{\rm AP}})$ by~\eqref{eq:ap_hot_bound}. Replacing its synthetic fields by the kinetic moments introduces only $\mathcal O(\varepsilon^{1+\sigma_{\rm AP}})$ in the momentum balance and $\mathcal O(\varepsilon^{1+\sigma_{\rm AP}})$ in the energy balance. These are $o(\varepsilon)$, so the viscous GK term is preserved. Finally, in~\eqref{eq:ap_gk_scaling}, divide the synthetic momentum equation by $\varepsilon$. Its forcing tends to zero, $\lambda/\varepsilon\to\bar n$, and the error in the scaled kinetic temperature is $\mathcal O(\varepsilon^{\sigma_{\rm AP}})$ by~\eqref{eq:ap_matching}. This proves~\eqref{eq:ap_gk_limit} for regular limits of both sets of fields.
\end{proofof}

\subsection{Ballistic limit}\label{app:ballistic}
Let $\delta_R,\delta_N\to0$, so that $\delta_C\to0$ and $\Kn_C\to\infty$. Along a ray of in-plane length $d$ the optical depth $\delta_Cd$ vanishes, and the limiting equation is free streaming, $\sin\phi\,\bm s\cdot\nabla I=0$: the incoming intensity is carried unchanged along each ray. This limit is not stiff, and Proposition~\ref{prop:ap} does not apply. The question is instead whether the synthetic coupling, whose response to representation errors grows like $\Kn_C$ (Section~\ref{sec:cis_fourier}), degrades free streaming.

\begin{proposition}[Ballistic limit]\label{prop:ballistic}
Let $I$ solve the frozen-source transport problem with source $\bm U_s=(T_s,\bm q_s)$ and incoming data $I_b$, and let $I_{\rm fs}(\bm x,\theta)=I_b(\bm x_b)$ be the free-streaming solution with the same incoming data. Then at every $(\bm x,\theta,\phi)$
\begin{equation}
 |I-I_{\rm fs}|\leq\Bigl[|I_b|+\frac{\|T_s\|_\infty+3\chi\|\bm q_s\|_\infty}{4\pi}\Bigr]\min\Bigl(1,\frac{\delta_Cd}{\sin\phi}\Bigr),
 \label{eq:ap_ballistic_pointwise}
\end{equation}
and, if every ray has in-plane length at most $d_{\max}$, the kinetic moments satisfy, componentwise,
\begin{equation}
 \|\mathcal K[I]-\mathcal K[I_{\rm fs}]\|_\infty\leq2\pi^2\delta_Cd_{\max}\Bigl[\|I_b\|_\infty+\frac{\|T_s\|_\infty+3\chi\|\bm q_s\|_\infty}{4\pi}\Bigr].
 \label{eq:ap_ballistic_moments}
\end{equation}
\end{proposition}
\begin{proofof}
By Eq.~\eqref{eq:characteristic}, $I-I_{\rm fs}=I_b(\mathrm{e}^{-\delta_Cd/\sin\phi}-1)+\int_0^d[A+\sin\phi\,B]\,\mathrm{e}^{-\delta_Ct/\sin\phi}\dd t/\sin\phi$. The first term and the $A$-term are bounded by $(|I_b|+\|T_s\|_\infty/(4\pi))(1-\mathrm{e}^{-\delta_Cd/\sin\phi})$, and, since $\delta_N=\chi\delta_C$, the $B$-term by $3\chi\|\bm q_s\|_\infty\min(\delta_Cd,\sin\phi)/(4\pi)$. Both factors are at most $\min(1,\delta_Cd/\sin\phi)$, which gives~\eqref{eq:ap_ballistic_pointwise}. The moments integrate against $\dd\Omega=\sin\phi\dd\phi\dd\theta$ with weights of modulus at most one, and $\int_0^\pi\min(1,\tau/\sin\phi)\sin\phi\dd\phi\leq\pi\tau$, which gives~\eqref{eq:ap_ballistic_moments}.
\end{proofof}

Three consequences follow. First, for every bounded source the kinetic moments approach those of free streaming at the rate $\delta_Cd_{\max}$; the synthetic source matters only through $\delta_C\bm U_s$. Second, at large $\Kn_C$ the synthetic map responds to a representation error with gain $\mathcal O(\Kn_C)$, $3k\Kn_C$ for the temperature in the Fourier setting (Section~\ref{sec:cis_fourier}). By~\eqref{eq:ap_ballistic_moments} that error reaches the kinetic moments with gain $\mathcal O(kd_{\max})$, independent of $\Kn_C$: the vanishing collision rate cancels the synthetic gain. Third, the separated state reproduces the limit with one polar mode. At $\delta_C=0$ every term of Eq.~\eqref{eq:polar} is proportional to $\sin\phi$, so the polar update returns a constant $\Phi_m$, and the spatial update~\eqref{eq:spatial} carries the inflow along $\bm s$. The corrections in~\eqref{eq:ap_ballistic_pointwise} are confined to grazing directions, $\sin\phi\lesssim\delta_Cd$, and account for the additional modes discussed in Section~\ref{sec:regimes_rank}.

For thermalizing walls $I_b$ is prescribed, and Eq.~\eqref{eq:ap_ballistic_moments} gives convergence to the free-molecular moments. On diffusely adiabatic walls the emission depends on the solution through Eq.~\eqref{eq:adiabatic}; Eq.~\eqref{eq:ap_ballistic_pointwise} then shows that $I$ satisfies the free-molecular wall problem up to $\mathcal O(\delta_Cd)$, and convergence of $I$ itself requires stability of that problem, which is not analyzed here. Periodic faces lengthen rays. Without a uniform $d_{\max}$, the factor $2\pi^2\delta_Cd_{\max}$ in~\eqref{eq:ap_ballistic_moments} is replaced by $\pi\int_0^{2\pi}\min(2/\pi,\delta_Cd)\dd\theta$, which is $\mathcal O(\delta_C\ln\delta_C^{-1})$ in cross-plane conduction. The proposition concerns the limit, not the convergence of the iteration; in an unbounded periodic medium the latter slows as $\rho(A_{\mathcal G})\to4/5$ (Section~\ref{sec:cis_fourier}).

\subsection{Stopping criteria, parameters and scope}\label{app:ap_scope}
For the diffusive and hydrodynamic limits, a sufficient accuracy prescription is
\begin{equation}
 \eta_n+\zeta_n+\omega_n\leq C_{\rm AP}\varepsilon^{1+\sigma_{\rm AP}},\qquad 0<\sigma_{\rm AP}\leq1;
 \label{eq:ap_stopping}
\end{equation}
choosing $\sigma_{\rm AP}=1$ gives $\mathcal O(\varepsilon^2)$ matching. Merely obtaining an absolute distribution error $o(1)$ does not preserve the leading $\mathcal O(\varepsilon)$ Fourier flux. The amplitude tests in Section~\ref{sec:stopping} do not control~\eqref{eq:ap_indicators}; their fixed $10^{-3}$ tolerances and finite caps are therefore empirical stopping rules, not a certificate of Proposition~\ref{prop:ap} for arbitrarily small Knudsen numbers. A joint update can reduce stale-mode errors, but a joint polar update alone does not enforce all spatial residual rows. Reaching~\eqref{eq:ap_stopping} uniformly with a prescribed rank and pair budget remains to be established.

For the parametric construction the proposition applies node by node if its bounds and indicators hold uniformly at those nodes. The parameter-averaged spatial equation~\eqref{eq:param_spatial} does not imply these pointwise bounds; a small measure-weighted residual can hide an inaccurate small-Knudsen node. Uniform certification requires pointwise control of the indicators, including the parameter-dependent polar approximation.

The estimates require spatial regularity, not only bounded amplitudes: derivatives of an unresolved layer may scale as $1/\varepsilon$. They also exclude the unsuppressed half-range wall channel of Section~\ref{sec:cis_fourier}. A global AP theorem needs consistent limiting wall conditions and stability of the macroscopic boundary problem. A fully discrete theorem additionally needs compatible DG/HDG moment operators, quadrature that preserves the collision invariants and degree-three cancellations, and uniform stability of the resulting projected solves. None follows solely from low-rank representability or a small full-order Fourier factor. The $\mathcal O(\varepsilon^2)$ diffusive contraction margin of source iteration, which the spatial sweeps of pure PGD inherit (Section~\ref{sec:cis_fourier}), explains the large iteration requirement of kinetic refresh~\cite{liu2022fast,adams2002fast}, but asymptotic acceleration and preservation of the limiting constitutive law are distinct properties.

\section{Fourier form of the alternating iteration}\label{app:fourier}
In a periodic homogeneous medium with an error mode $\mathrm{e}^{ikx}$, write $X_r=\mathrm{Re}(\hat x_r(\theta)\mathrm{e}^{ikx})$ and $(T_s,\bm q_s)=\mathrm{Re}((\hat T,\hat{\bm q})\mathrm{e}^{ikx})$. Since $\int\mathrm{Re}(a\mathrm{e}^{ikx})\mathrm{Re}(b\mathrm{e}^{ikx})\dd x=\frac L2\mathrm{Re}(a\bar b)$ over a period, the coefficients of the PGD--GSIS polar update~\eqref{eq:polar} become, after dropping the common factor $L/2$,
\begin{equation}
\begin{gathered}
 \alpha_{mr}=\mathrm{Re}\!\int\!\hat x_m\overline{\hat x_r}\dd\theta,\qquad
 \beta_{mr}=\mathrm{Re}\!\int\!\hat x_m\overline{ik\cos\theta\,\hat x_r}\dd\theta,\\
 F_{T,m}=\frac{\delta_C}{4\pi}\mathrm{Re}\!\int\!\hat x_m\overline{\hat T}\dd\theta,\qquad
 F_{q,m}=\frac{3\delta_N}{4\pi}\mathrm{Re}\!\int\!\hat x_m\overline{\bm s\cdot\hat{\bm q}}\dd\theta,
\end{gathered}
\end{equation}
with $\beta_{mm}=0$ and $\mu_m=0$, and the spatial update~\eqref{eq:spatial} is solved exactly:
\begin{equation}
 \hat x_m(\theta)=\frac{c_m\hat T+d_m\,\bm s\cdot\hat{\bm q}-\sum_{r<m}(a_{mr}+ikb_{mr}\cos\theta)\hat x_r(\theta)}{a_{mm}+ikb_{mm}\cos\theta}.
\end{equation}
Each new spatial factor starts constant in $\theta$, as in the solver, and the iteration uses the quadrature, caps and normalization of Section~\ref{sec:results}; the initial error is $\bm U^0=(1,0.5i,0.5)$. In our tests, random initial errors left $\gamma_{\mathcal G}$ and $n_{12}$ essentially unchanged and changed $\|e\|$ by at most a factor of about $2.5$. The kinetic map is Eq.~\eqref{eq:macro_reconstruction}. For the synthetic map, the third moments of the separated state need only $\mathsf M_2$ and $\mathsf M_4$,
\begin{equation}
\begin{aligned}
 M_{xxx}&=\sum_r\int\hat x_r\Bigl[\cos^3\theta\,\mathsf M_4(\Phi_r)-\tfrac35\cos\theta\,\mathsf M_2(\Phi_r)\Bigr]\dd\theta,\\
 M_{yxx}&=\sum_r\int\hat x_r\Bigl[\cos^2\theta\sin\theta\,\mathsf M_4(\Phi_r)-\tfrac15\sin\theta\,\mathsf M_2(\Phi_r)\Bigr]\dd\theta,
\end{aligned}
\end{equation}
and substituting $\nabla\to ik\bm e_x$ into Eq.~\eqref{eq:synthetic} with zero data gives $\hat q_x=0$ from the energy equation and Eq.~\eqref{eq:G_fourier} from the two momentum equations.

The full-order operators $A_{\mathcal K}$ and $A_{\mathcal G}$ are obtained by applying the two maps to the exact response $I_{\rm ex}=(\delta_C\hat T+3\delta_N\bm\Omega\cdot\hat{\bm q})/[4\pi(\delta_C+ik\sin\phi\cos\theta)]$ on the same grid. $A_{\mathcal K}$ is the full-order source iteration analyzed in~\cite{liu2022fast}; its spectral radius tends to one as $k\Kn_C\to0$. In the opposite limit, for a temperature error, $M_{xxx}=\hat T\cdot\frac12\int_{-1}^1(\xi^3-\frac35\xi)(1+i\hat k\xi)^{-1}\dd\xi\to-4\hat T/(15i\hat k)$ as $\hat k=k\Kn_C\to\infty$, where $\xi=\Omega_x$. Equation~\eqref{eq:G_fourier} then gives $\hat T'\to-\frac45\hat T$, while the flux channels vanish at this order. Hence $\rho(A_{\mathcal G})\to4/5$, whereas $\rho(A_{\mathcal K})=\mathcal O(1/\hat k)$.

\paragraph{Spatial sweep of pure PGD} For a fixed polar factor $\Phi$, write $a=\delta_C\int\Phi^2\dd\phi$, $b=\int\sin\phi\,\Phi^2\dd\phi$, $c\mathsf T=\frac{\delta_C}{4\pi}\int\Phi\dd\phi\,\mathsf M_1(\Phi)$ and $d\mathsf Q=\frac{3\delta_N}{4\pi}\mathsf M_1(\Phi)\mathsf M_2(\Phi)$, the coefficients of Eqs.~\eqref{eq:spatial}--\eqref{eq:wcoeff} with the current-mode moments on the right. One lagged sweep maps the error $\hat x$ of the current mode to
\begin{equation}
 \hat x'(\theta)=\frac{c\mathsf T\,\hat Y+d\mathsf Q\,\bm s(\theta)\cdot\hat{\bm Z}}{a+ikb\cos\theta},\qquad
 \hat Y=\int_0^{2\pi}\hat x\dd\theta,\quad \hat{\bm Z}=\int_0^{2\pi}\bm s\,\hat x\dd\theta,
 \label{eq:sweep_fourier}
\end{equation}
which is a linear map of $(\hat Y,\hat Z_x,\hat Z_y)$; $\rho_{\rm sw}$ in Table~\ref{tab:fourier} is the spectral radius of its $3\times3$ matrix, evaluated with the quadrature of Section~\ref{sec:results}. For $\Phi=1$, the effective extinction is $a/b=\pi\delta_C/2$, the isotropic in-scattering ratio $2\pi c\mathsf T/a$ equals one, and the flux ratio $\pi d\mathsf Q/a$ equals $3\chi/4$. For $\Phi=\sin\phi$, $a/b=3\pi\delta_C/8$, the isotropic ratio is again one, and the flux ratio equals $\chi$. As $k/\delta_C\to0$, the sweep therefore approaches the identity on the conserved temperature channel, and also on the flux channel of $\Phi=\sin\phi$ as $\chi\to1$. In the same periodic setting $\beta_{mm}=0$, and an azimuthally isotropic $\hat x_m$ gives $2\alpha^T_{mm}=\alpha_{mm}$ and $\alpha^q_{mm}=0$. The operator on the left of Eq.~\eqref{eq:polar_implicit} then maps every $\Phi$ to a function with zero $\sin\phi$-moment, so the polar update of pure PGD is singular without the wall and jump terms of $\beta_{mm}$. For this reason we analyze the spatial sweep of pure PGD rather than executing its iteration in Fourier form.

\section{DG streaming form}\label{app:dg}
Let $V_h$ be the piecewise polynomials of degree $p$ on the triangles $K$~\cite{reed1973triangle,cockburn2001runge}, and $a^\pm=(a\pm|a|)/2$. With prescribed inflow excluded,
\begin{equation}
 b_h(V,U)=\sum_K\int_0^{2\pi}\Bigl[-(\bm s\cdot\nabla V,U)_K
 +\langle V,(\bm s\cdot\bm n)^+U\rangle_{\partial K}
 +\langle V,(\bm s\cdot\bm n)^-U^{\rm ext}\rangle_{\partial K\cap\mathcal F_i}\Bigr]\dd\theta,
 \label{eq:bh}
\end{equation}
where $\mathcal F_i$ collects interior and periodic faces and $U^{\rm ext}$ is the neighboring (or partner) trace. Taking $V=U$ gives $b_h(U,U)=\frac12\int_0^{2\pi}\bigl[\sum_{F\in\mathcal F_i}\int_F|\bm s\cdot\bm n|\,[\![U]\!]^2+\int_{\partial\mathcal D_w}|\bm s\cdot\bm n|U^2\bigr]\dd\theta\geq0$, with $\partial\mathcal D_w$ the wall faces; this is used in Section~\ref{sec:factor_eqs}. The thermal-wall load is $g_b(V)=\int_{\partial\mathcal D}\int_0^{2\pi}V(\bm s\cdot\bm n)^-T_b/(4\pi)\dd\theta\dd S$, so that $\beta_{mr}=b_h(X_m,X_r)$ and $\mu_m=g_b(X_m)$ in Eq.~\eqref{eq:polar}; the adiabatic load replaces $T_b/(4\pi)$ by the wall emission~\eqref{eq:adiabatic}, and periodic faces, which belong to $\mathcal F_i$, contribute only the offset $\Delta T_b/(4\pi)$ as a load. For an azimuth $\theta_j$ and $X_r|_K=\sum_lx^{K,j}_{r,l}\varphi_l$, Eq.~\eqref{eq:spatial} becomes
\begin{equation}
\begin{split}
 (a_{mm}M_K+b_{mm}B_{K,j})\bm x_m^{K,j}
 ={}&M_K(c_m\bm T_{s,K}+d_m\bm s_j\cdot\bm q_{s,K})-\frac{\mathsf T_m}{4\pi}\bm f^b_{K,j}
 -b_{mm}\sum_{F}C_{KF,j}\bm x_m^{{\rm ext},j}\\
 &-\sum_{r=1}^{m-1}\Bigl[(a_{mr}M_K+b_{mr}B_{K,j})\bm x_r^{K,j}+b_{mr}\sum_FC_{KF,j}\bm x_r^{{\rm ext},j}\Bigr],
\end{split}\label{eq:dg_matrix}
\end{equation}
written for PGD--GSIS (pure PGD replaces $\bm T_{s,K}$ and $\bm q_{s,K}$ by the kinetic moments, with the current-mode part from the previous sweep), with $M_K$ the mass matrix, $B_{K,j}$ the volume and outflow terms, $C_{KF,j}$ the coupling to the upwind neighbor across an inflow face $F$, and $\bm f^b_{K,j}$ the thermal inflow load with entries $\langle\varphi_l,(\bm s_j\cdot\bm n)^-T_b\rangle_{\partial K\cap\partial\mathcal D}$. The parametric form uses $\bar a$, $\bar b$, $S_{T,m}$ and $\bm S_{q,m}$.

\section{HDG discretization of the synthetic system}\label{app:hdg}
\paragraph{Mixed form} With the auxiliary tensor $\bm L$ and the viscous flux $\bm\Pi$,
\begin{equation}
 \Pi_{xx}=\tfrac43L_{xx}-\tfrac23L_{yy},\qquad
 \Pi_{xy}=\Pi_{yx}=L_{xy}+L_{yx},\qquad
 \Pi_{yy}=\tfrac43L_{yy}-\tfrac23L_{xx},
 \label{eq:sigma}
\end{equation}
Eq.~\eqref{eq:synthetic} is written on each element as
\begin{equation}
 \nabla\cdot\bm q=0,\qquad
 \frac13\nabla T+\delta_R\bm q-\nabla\cdot\bm\Pi=0,\qquad
 \bm L-\frac{\Kn_C}{5}\nabla\bm q=\Kn_C\,\mathsf H,
 \label{eq:mixed}
\end{equation}
where $L_{ij}$ approximates $\frac{\Kn_C}{5}\partial_jq_i$ and $\mathsf H=(H_{xx}+\frac12H_{yy},\ \frac12H_{xy},\ \frac12H_{xy},\ \frac12H_{xx}+H_{yy})$ in the ordering $(L_{xx},L_{xy},L_{yx},L_{yy})$. With $\partial_z=0$, Eq.~\eqref{eq:sigma} reproduces $\frac{\Kn_C}{5}[\nabla^2\bm q+\frac13\nabla(\nabla\cdot\bm q)]$, and the chosen $\mathsf H$ gives $\bm\Pi(\mathsf H)=\bm H$, so that $\nabla\cdot\bm\Pi$ contains exactly $\Kn_C\nabla\cdot\bm H$. The element unknowns are $\bm u_K=(T,q_x,q_y,L_{xx},L_{xy},L_{yx},L_{yy})\in[P_p(K)]^7$ and the face unknowns $\hat{\bm u}_F=(\hat T,\hat q_x,\hat q_y)\in[P_p(F)]^3$.

\paragraph{Local problem} With stabilization $\tau_T=\tau_q=1$ and test function $\varphi\in P_p(K)$,
\begin{subequations}\label{eq:hdg_local}
\begin{align}
 -(\bm q,\nabla\varphi)_K+\langle\bm q\cdot\bm n+\tau_TT,\varphi\rangle_{\partial K}
 &=\langle\tau_T\hat T,\varphi\rangle_{\partial K},\\
 \delta_R(q_i,\varphi)_K-\tfrac13(T,\partial_i\varphi)_K-(\varphi,\partial_k\Pi_{ik})_K+\langle\tau_qq_i,\varphi\rangle_{\partial K}
 &=\langle\tau_q\hat q_i-\tfrac13n_i\hat T,\varphi\rangle_{\partial K},\\
 (L_{ij},\varphi)_K+\frac{\Kn_C}{5}(q_i,\partial_j\varphi)_K
 &=\frac{\Kn_C}{5}\langle n_j\hat q_i,\varphi\rangle_{\partial K}+\Kn_C(\mathsf H_{ij},\varphi)_K,
 \label{eq:hdg_L}
\end{align}
\end{subequations}
where $\mathsf H_{ij}$ is built from the weak load~\eqref{eq:weak_hot}. In matrix form $A_K\bm u_K=B_K\hat{\bm u}_{\partial K}+\bm f_K[I_m]$. The matrix $A_K$ depends on the mesh, $\delta_R$ and $\Kn_C$ only; it is factorized once per element (and per parameter node), and $A_K^{-1}B_K$ is stored.

\paragraph{Global problem} Single-valuedness of the numerical energy flux $\bm q\cdot\bm n+\tau_T(T-\hat T)$ and of the momentum flux $\frac13\hat T\bm n-\bm\Pi\bm n+\tau_q(\bm q-\hat{\bm q})$ on interior faces, after elimination of $\bm u_K$, gives a sparse unsymmetric system $\mathbb K\hat{\bm U}=\bm F[I_m]$ for the traces. It is factorized once by PARDISO~\cite{schenk2004solving}; every synthetic refresh reassembles $\bm F$ and performs a backsolve. Periodic faces are identified with their partners, and the offset $\Delta T_b$ enters the energy component of $\bm F$. Without a thermalizing wall the system is bordered by
\begin{equation}
 \frac{1}{|\mathcal D|}\int_{\mathcal D}T\dd\bm x=T_\ast
 \label{eq:gauge}
\end{equation}
through a Lagrange multiplier.

\paragraph{Wall closure} On a physical face the conservation condition is replaced by a linearized kinetic wall law~\cite{Su2020JCP,liu2022fast}. Let $\bm U=(T,q_x,q_y)$ be the trace, $\bm U^{\kin}$ the corresponding moments of $I_m$, and $\bm\psi(\bm\Omega)=(\bm\Omega\cdot\bm n)(1,\Omega_x,\Omega_y)$. The outgoing half-range carries the equilibrium $I^{\rm eq}(\bm U)=\bm a(\bm\Omega)\cdot\bm U$, $\bm a=(1,3\chi\Omega_x,3\chi\Omega_y)/(4\pi)$, whose flux derivative is $J_F^+=\int_{\bm\Omega\cdot\bm n>0}\bm\psi\,\bm a^\top\dd\Omega$. On a thermalizing wall the incoming state is fixed and $J_F=J_F^+$. On a diffuse adiabatic wall the incoming level~\eqref{eq:adiabatic} is linear in $\bm U$, which adds the rank-one term $J_F^-=\bigl(\int_{\bm\Omega\cdot\bm n<0}\bm\psi\dd\Omega\bigr)\bm e_+^\top/(-D)$, where $\bm e_+^\top$ is the first row of $J_F^+$ and $D=\int_{\bm\Omega\cdot\bm n<0}\bm\Omega\cdot\bm n\dd\Omega$. The imposed law is the affine expansion
\begin{equation}
 \widehat{\bm F}(\bm U)=\bm{\mathcal F}^{\kin}[I_m]+J_F(\bm U-\bm U^{\kin}),\qquad
 \bm{\mathcal F}^{\kin}[I_m]=\int\bm\psi(\bm\Omega)I_m^{\rm up}\dd\Omega,
 \label{eq:wall_affine}
\end{equation}
with $I_m^{\rm up}$ the upwind state of the kinetic sweep. At $\bm U=\bm U^{\kin}$ it returns the kinetic wall flux of $I_m$. All half-range integrals use the discrete ordinates of the kinetic solver, and $J_F$ is computed once per wall face (and parameter node). One synthetic refresh consists of assembling $\mathsf H$ and $\bm{\mathcal F}^{\kin}$ from $I_m$, backsolving the local particular solutions, solving for the traces, and recovering $T^{\syn}$ and $\bm q^{\syn}$ element by element (after the shift~\eqref{eq:gauge} if needed).

\bibliographystyle{elsarticle-num}
\bibliography{ref}
\end{document}